\documentclass{ws-ijmpe}
\usepackage[super,compress]{cite}
\usepackage{dcolumn}
\usepackage{bm}
\usepackage{amsfonts}
\usepackage{setspace}
\usepackage{float}

\newcommand{\be}{\begin{equation}}
\newcommand{\ee}{\end{equation}}
\newcommand{\ba}{\begin{eqnarray}}
\newcommand{\ea}{\end{eqnarray}}
\newcommand{\bd}{\begin{displaymath}}
\newcommand{\ed}{\end{displaymath}}
\renewcommand{\vec}[1]{\mbox{\boldmath$#1$}}

\begin{document}

\markboth{L.P. Csernai, S. Velle}{Study of Rotating High Energy Systems with the Differential HBT Method}

\catchline{}{}{}{}{}

\title{Study of Rotating High Energy Systems with the Differential HBT Method}

\author{L.P. Csernai}

\address{Institute of Physics and Technology, University of Bergen, 
Allegaten 55, 5007 Bergen, Norway 
csernai@ift.uib.no}

\author{S. Velle}

\address{Institute of Physics and Technology, University of Bergen, 
Allegaten 55, 5007 Bergen, Norway 
sindre.velle@ift.uib.no}

\maketitle

\begin{history}
\received{Day Month Year}
\revised{Day Month Year}
\end{history}

\begin{abstract}
Peripheral heavy ion reactions at ultra relativistic energies 
have large angular momentum that can be studied via two particle 
correlations using the Differential Hanbury Brown and Twiss method. 
In the present work we analyze the possibilities and sensitivity of 
the method in rotating, few source systems. 
Analytic results provide insight in the advantages of this method.
\end{abstract}

\keywords{Two particle correlation; Rotation; Heavy ion collisions.}

\ccode{PACS numbers: 25.75.Gz, 25.75.Nq}


\section{Introduction}
\label{I}

Collective flow is one of the most dominant observable features
in heavy ion reactions up to the highest available energies, and
its global symmetries as well as its fluctuations are extensively
studied. Especially at the highest energies for peripheral reaction
the angular momentum of the initial state is substantial, which leads
to observable rotation according to fluid dynamical estimates
\cite{hydro1}.
Furthermore the low viscosity quark-gluon fluid may lead to
to initial turbulent instabilities, like the 
Kelvin Helmholtz Instability (KHI), according to numerical
fluid dynamical estimates
\cite{hydro2}, 
which is also confirmed in a simplified analytic model
\cite{KHI-Wang}. 
These turbulent phenomena further increase
the rotation of the system, which also leads to a large 
vorticity and circulation of the participant zone one order of 
magnitude larger than from random fluctuations 
in the transverse plane \cite{CMW12, FW11-1,FW11-2}.
It is estimated \cite{hydro2} that the increased rotation 
can be observable via the increased $v_1$-flow, but the $v_1$
signal at high energies is weak, so other observables of the
rotation are also needed.

The two particle correlation method is used to determine the
space-time size of the system emitting the observed particles,
thus providing valuable information on the exploding and expanding
system at the freeze out stage of a heavy ion collision. This method
is based on the Hanbury Brown and Twiss (HBT) method, originally used 
for the determination of the size of distant stars 
\cite{StarHBT-1,StarHBT-2}.
In heavy ion collisions the HBT method was used first for the
same purpose, the determination of the system size 
\cite{FirstHBTs},
but later also the ellipsoidal shape of the system and its tilt
\cite{DM95,MLisa-1,MLisa-2,MLisa-3}.
It was also observed relatively early that the expansion of the
system modifies the size estimates due to the collective radial
flow velocity of the emitting system
\cite{Pratt}, while the effect of flow on two particle correlations
was also analysed in great detail
\cite{Sin89}.
Transport model studies have indicated that the HBT radius
shows a minimum at the phase transition threshold
\cite{QfLi07-9-1,QfLi07-9-2}.

Recently in AdS/CFT holography we have seen \cite{McInnes} that 
the region of the phase diagram is dependent on the amount of 
angular momentum as well as instabilities in the plasma produced in 
these peripheral collisions. Whether the angular momentum is carried 
by rotation or by shear is of importance.

In heavy ion reactions the angular momentum is expected to 
generate local vorticity \cite{CMW12},
which equilibrates with the 
internal angular momentum of the constituted particles and causes 
significant observable polarization after the final particle 
emission \cite{BCW2013}.

In Ref.~\refcite{CVW2014} a method is used to do fluid dynamical model calculations 
and it was demonstrated that the introduced Differential HBT method can 
be used to measure the rotation of the system. We study this method hereby 
in simple and transparent few source fluid dynamical model with different
symmetry structures. These type of models are frequently used to 
demonstrate different effect in a transparent way
\cite{Eyyubova-1,Eyyubova-2, Vovchenko, Zschocke}.
These studies provide analytic results, show how we can detect
rotation via two particle correlation functions, and what
effects may cause difficulties in identifying rotation.

\subsection{The Emission Function}
\label{TEF}

The emission function and the hydrodynamical parametrization together with 
the Freeze Out (FO) layer have been studied \cite{CVW2014} earlier. 
The emission probability is proportional to \ $G(x) \, H(\tau)$:
$$
S(x,\vec p) d^4 x = 
  p^\mu\  \hat\sigma_\mu(x) \ G(x)\, H(\tau)\, \ d\tau d^3x \  f(x,p) \ ,
$$
where 
\begin{equation}
H(\tau) = \frac{1}{(2\pi \Theta^2 )^{1/2}} 
\exp\left[-\frac{(\tau-\bar{\tau})^2}{2 \Theta^2 }\right] ,
\label{S-BL2}
\end{equation}
and $G(x)$ is the ST emission density across the layer of the particles. 
For the phase space distribution we frequently use the 
J\"uttner (relativistic Boltzmann) distribution, in terms of the local invariant
scalar particle density the J\"uttner distribution is \cite{Juttner}
\be
f^J(x,p) = \frac{n(x)}{C_n} 
\exp\left(-\frac{p^\mu u_\mu (x)}{T(x)}\right)\ ,
\label{Jut-2}
\ee
where $C_n = 4 \pi m^2 T K_2(m/T)$.

\subsection{Pion Correlation Functions}
\label{PCF}

The pion correlation function is defined as the inclusive two-particle 
distribution divided by the product of the inclusive one-particle 
distributions, such that \cite{WF10}:
\begin{equation}
C( p_1 , p_2) = 
\frac{P_2( p_1, p_2)}{P_1( p_1)P_1( p_2)},
\end{equation}
where $ p_1$ and $ p_2$ are the 4-momenta of the pions 
and \vec{k} and \vec{q} are the average and relative momentum respectively. 

We use a method for
moving sources presented in Ref.~\refcite{Sinyukov-1}. 
In the formulae the $\hbar = 1$ convention is used and $k$ and $q$ are
considered as the wavenumber vectors.
The correlation function is:
\begin{equation}
C(k,q) =
1 + \frac{R(k,q)}{\left| \int d^4 x\,  S(x, k) \right|^2}\ ,
\label{C-def}
\end{equation}
where
\be
R(k,q) = \int d^4 x_1\, d^4 x_2\, \cos[q(x_1-x_2)]  
S(x_1,  {k}+{q}/2)\, S(x_2,{k}-{q}/2)\ .
\label{R1}
\ee
Here $R(k,q)$ can be calculated \cite{Sinyukov-1} via the function
\be
 J(k,q) =\! \int\! d^4x\ S(x,k+q/2)\, \exp(iqx) =\! 
          \int\! d^4x\ S(x,k+q/2)\, [\cos(qx) + i \sin(qx)] \ ,
\label{J-def}
\ee
and we obtain the $R(k,q)$ function as
\be
R(k,q) = {\rm Re}\, [ J(k,q)\, J(k,-q) ]
\label{R-def}
\ee
Thus, the expression of the correlation function, Eq. (\ref{R1}) will
be modified to
\be
R(k,q) = \int\!  d^4 x_1 d^4 x_2\, S(x_1, k) S(x_2, k) 
    \cos[q(x_1{-}x_2)] 
    \exp\left[ -\frac{q}{2} \cdot \left( \frac{u(x_1)}{T(x_1)}
       -\frac{u(x_2)}{T(x_2)} \right)\right],
\label{R-def2}
\ee
and the corresponding $J(k,q)$ function will become
\be
J(k,q) =  \int d^4x\ S(x,k)\, 
\exp\left[ - \frac{q \cdot u(x)}{2T(x)} \right]\, \exp(iqx)\ .
\label{J2} 
\ee

\subsection{One Fluid Cell as Source}
\label{SSofcs}

We now assume a source function, which is reduced to one Freeze Out (FO)
time moment. Thus the integration over the 4-volume of an emission
layer is reduced to the 3-volume of a FO hypersurface
\cite{CF}. For simplicity,
we assume FO along the timelike coordinate, $t$, where we assume a
local J\"uttner distribution. Thus, we
have the source function as 
\begin{equation}
S({x}, k) =
G(x)\,H(t) 
\exp\left(-\frac{k_\mu u^\mu (x)}{T(x)} \right)
k^\mu\, \hat\sigma_\mu\ ,
\end{equation}
where $k^\mu \hat\sigma_\mu $ is an invariant scalar and 
$ \hat\sigma_\mu $ is the direction of emission 
unit vector \cite{Cso-5, CVW2014}, and for
a single cell we use a simple quadratic parametrization for $n(x)$ as:
\begin{equation}
G(x) = \gamma n(x) = 
\gamma n_s  \exp\left( - \frac{x^2 + y^2 + z^2}{2 R^2} \right) .
\end{equation}
Here $n_s$ is the average density of the Gaussian source, $s$, (or fluid cell)
of mean radius $R$.

\bigskip
{\bf Single moving source:}
\noindent
Let us take a single source which moves
in the x-direction with a velocity $v_x$. Then we have,
$  u_s^\mu = \gamma_s (1,v_x,0,0) $, and the scalar product
$k\cdot u_s/T_s = k_\mu u_s^\mu / T_s $ provides an additional
contribution to the correlation function. However, in the case
of a single fluid cell or a single source the
velocity and the temperature do not change within
the cell, so the modifying term in
eq. (\ref{R-def2}) becomes unity.
The source function becomes
\begin{equation}
S({x}, k) =
\frac{n(x) \, (k^\mu\,\hat\sigma_\mu)}{C_n} \, 
\exp\left[-\frac{k \cdot u_s}{T_s}\right] \ ,
\end{equation}
where $[k\cdot u_s/T_s ] = [\gamma_s (E_k - k_x v_x)/T_s]$ 
as used in Ref.~\refcite{Cso-5,CVW2014}.

Within the single source (or fluid element) the velocity $u_s$ and temperature
$T_s$ are assumed to be the same. The source or fluid element
may have a density profile, but this profile should be the same
for all cells (although the average density, $n_s$ must not be the same. 
The spatial integrals can be performed in the
rest frame of the cell, giving the same integral.
In this simplest case we also assume that the FO direction is
$\hat\sigma^\mu = (1, 0, 0, 0)$, so the $\tau$-coordinate coincides with
the $t$-coordinate, and it is orthogonal to the $x, y, x-$ coordinates.

We can perform the integral along the $t$ direction of
$H(t)$, which gives unity and then
the single particle distribution is
\be
\begin{split}
&  \int\! d^4x\ S(x, k) = \frac{n_s (k^\mu \hat\sigma_\mu) }{C_n}
        \exp\left(-\frac{k \cdot u_s}{T_s}\right) \times \\ &
  \int_{-\infty}^{+\infty} \!\!\!\!\!\!\!\!
                        H(t)  dt 
  \int_{-\infty}^{+\infty} \!\!\!\!
                        e^{-\frac{x^2}{2R^2} } dx 
  \int_{-\infty}^{+\infty} \!\!\!\!
                        e^{-\frac{y^2}{2R^2} } dy 
   \int_{-\infty}^{+\infty} \!\!\!\! 
                        e^{-\frac{z^2}{2R^2} } dz \\  \
& = n_s \ (k^\mu \hat\sigma_\mu) \ \exp\left(-\frac{k \cdot u_s}{T_s}\right)
    \frac{\left(2 \pi R^2 \right)^{3/2} }{C_n} \ ,
\end{split}
\label{InS}
\ee
where $T_s$ is the temperature of the source. The contribution to the nominator from Eq. (\ref{J2}) is
\be
\begin{split}
& 
J(k,q) = \int  d^4x\,  e^{i  q \cdot  x} e^{-q^0/(2T_s)} S({x}, k) \exp\left[ - \frac{q \cdot u_s}{2T_s}\right]  \  = \\
& \frac{n_s (k^\mu \hat\sigma_\mu)}{C_n} \left(2 \pi R^2 \right)^{3/2} 
 \exp\left[{-}\frac{k^0}{T_s}\right] \exp\left[{-}\frac{q^0}{2T_s}\right] \times \\
& \exp\left[-\frac{R^2}{2} q^2\right] 
\exp\left[-\frac{\Theta^2}{2} (\hat\sigma^\mu q_\mu)^2\right] 
\exp\left[ - \frac{q \cdot u_s}{2T_s} \right] \ ,
\end{split}
\ee
In the time integral the present choice of $\hat\sigma^\mu$ would
give $(q^0)^2$, but we wanted to indicate that other choices are 
also possible and they would yield $(\hat\sigma^\mu q_\mu)^2$.
In the $J(k,q) J(k,-q)$ product the terms $\exp[\pm q^0 /(2T_s)]$
cancel each other
Also in the $J(k,q) J(k,-q)$ product 
the terms $\exp[\pm q \cdot u_s /(2T_s)]$ cancel each other, 
so the correlation function will not be dependent on the velocity in this case.
Inserting these equations into (\ref{C-def}) we get
\begin{equation}
C( k,  q) = 1 +
\exp\left(-(\Delta \tau)^2 (\hat\sigma^\mu q_\mu)^2 - R^2 q^2\right) \ .
\label{Csst}
\end{equation}
If we have a source at a point in the FO layer, which is at a longer 
distance from the external side of the FO layer than $\Theta$, then
the contribution of the time integral from this point is reduced.
In a few source model it is more transparent to describe this reduction 
by assigning a smaller weight factor to the contribution of the deeper 
lying source. See this in more detail in section \ref{ASm}.

If we tend to an infinitely narrow FO layer, 
$\Theta \rightarrow 0$, i.e. to a FO hypersurface, then
\vskip -3mm
\begin{equation}
C( k,  q) = 1 +
\exp\left(-R^2 q^2\right) \ .
\label{Csss}
\end{equation}
The $ k$ dependence drops out from the correlation function, 
$C( k,  q)$ as the $ k$ dependent parts are separable. 
The size of the fluid cells in a high resolution 3+1D fluid dynamical
calculation is $(0.3$fm$)^3$ \cite{CVW2014}. With this resolution the
{\it numerical viscosity} of the fluid dynamical calculation 
\cite{Horvat} is the same as the estimated minimal viscosity
of the QGP \cite{Kovtun} which occurs at the critical point
of the phase transition \cite{CsKM}. Eq. (\ref{Csss}) returns the 
standard correlation function for spherical source.
%

\subsection{Two moving sources}\label{Ltsm} 

\begin{figure}[ht] 
\begin{center}
      \includegraphics[width=6cm]{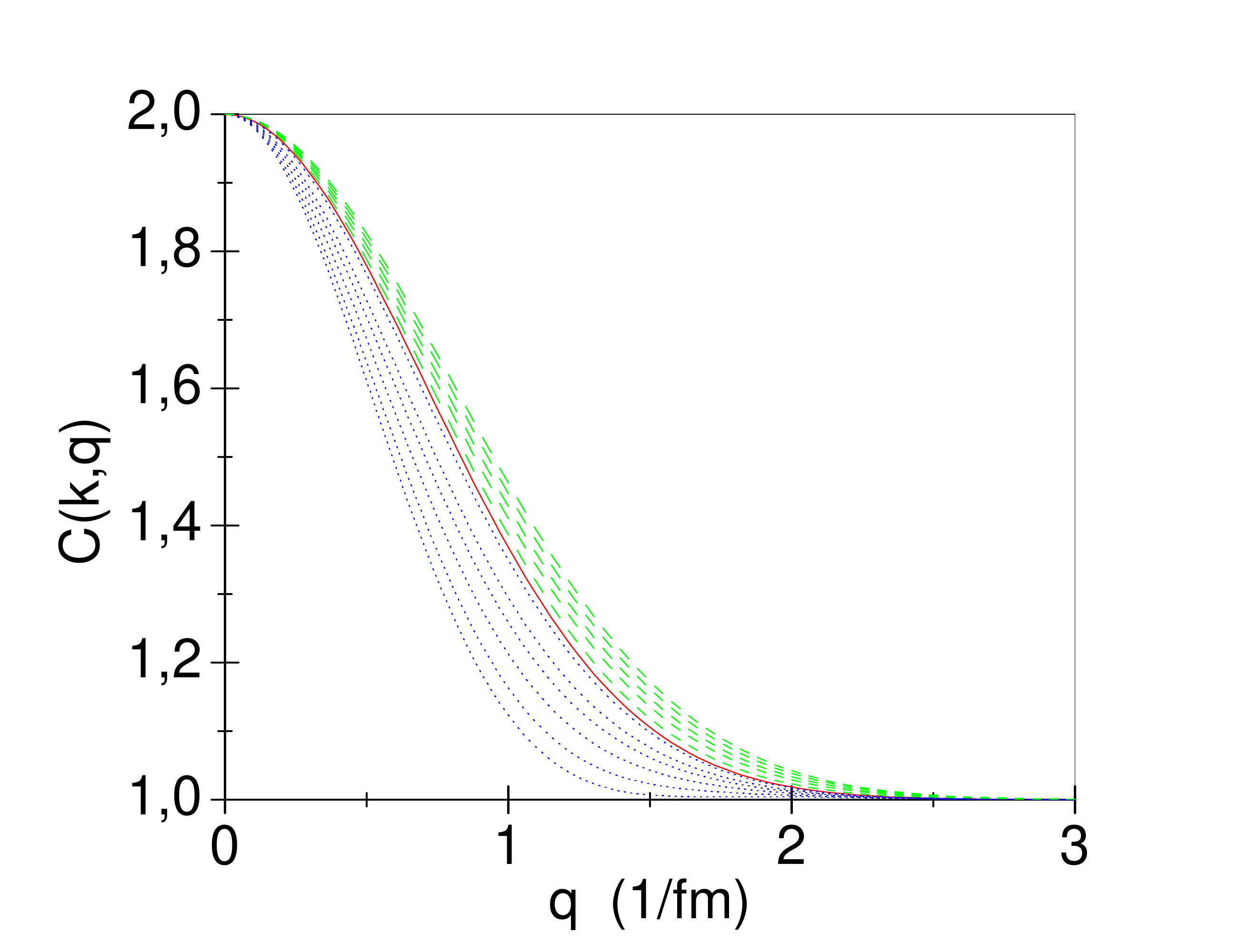}
\end{center}
\caption{ (color online)
Two moving sources in the reaction ($[x-z]$) plane with a distance
between them of $2d$ in the $z-$direction. The sources are moving in the
directions indicated by the (red) arrows.}
\label{F-1}
\end{figure}

For emission from two moving sources,
two particle correlations were studied in 
Ref.~\refcite{Cso-5}. Here we use the present method. Now we assume that 
the two source system is symmetric both their positions are placed
symmetrically and 
also their FO normal vectors, $\hat\sigma^\mu$, are the same (Fig. \ref{F-1}).
If the normal were $\hat\sigma^\mu =(1,0,0,0)$, then the
invariant scalar $k^\mu \hat\sigma_\mu$ would be $k^0 = E_k$,
although we do not need this additional requirement to illustrate
the correlation function, which would arise from an idealized symmetric 
system. 
We also assume that the time distributions, $H(\tau)$ for the
two sources are identical, so these can be integrated simultaneously 
and yield unity.  
We now have two sources moving in opposite directions, so that\\
$u_s = u_1$ or $u_2$ where
$u_1^\mu = (\gamma_s, \gamma_s \vec v_1)$, 
$u_2 = \bar u_s^\mu = (\gamma_s, \gamma_s (- \vec v_1))$,   and
$\vec u_s \equiv \gamma_s\, \vec v_s$,      so that
$\vec u_1 = - \vec u_2$, see Fig. \ref{F-1}. Similarly,
$x_s = x_1$ or $x_2$ where
$x_s^\mu = (t_s, \vec x_s)$, 
$\bar x_s^\mu = (t_s, - \vec x_s)$,
and $\vec x_1 = - \vec x_2$.  For now we also assume that
FO happens at a $t=$const. FO hypersurface, so 
$d\hat\sigma^\mu = (1, 0,0,0)$ and so $t_1 = t_2$.

\noindent
If we have 
several sources then the source function in J\"uttner approximation is
\be
S({x}, k) = \sum_s S_s(x,k) = 
(k^\mu\, \hat\sigma_\mu) \ \sum_s \frac{n_s(x) \, }{C_{ns}} \, 
\exp\left[-\frac{k \cdot u_s}{T_s}\right] \ ,
\ee
while the $J$ function is
\be
 J(k,q) =  \sum_s  
         \exp\left[- \frac{q \cdot u_s}{2T_s}\right] 
 \exp(iqx_s) \int_S d^3x\ S_s(x,k)\, \exp(iqx)\ ,
\ee
where $x_s$ is the 4-position of the center of source $s$, 
and the spatial integrals run separately for each of the 
identical sources, i.e. we assume fluid cells with identical density
profiles, but with different densities, $n_s$,
velocities, $u_s$ and temperatures, $T_s$.

The spatial integral 
for one source is the same as for a single source. Thus,
\be
\begin{split}
& \int\! d^3x\   S(x, k) = \sum_s \int_S d^3x\ S_s(x,k) = \\
  (k^\mu\, \hat\sigma_\mu) \ & \left(2 \pi R^2 \right)^{3/2} \
    \  \frac{n_s}{C_{ns}}  \exp\left(-\frac{k^0 \gamma_s}{T_s}\right) 
 \left[ 
\exp\left( \frac{\vec k \vec u_s}{T_s} \right)+ 
\exp\left(-\frac{\vec k \vec u_s}{T_s} \right) 
\right] \ .
\end{split}
\ee
The function $J(k,q)$ becomes
\be
\begin{split}
& J(k,q) = 
 \sum_s \exp\left[-\frac{q \cdot u_s}{2 T_s} \right]\, 
\exp(iqx_s)  \int_S d^4x\ S_s(x,k)\, \exp(iqx)  = \\
& 
     (k^\mu\, \hat\sigma_\mu) \ \left(2 \pi R^2 \right)^{3/2}  
\exp\left(-\frac{R^2 q^2}{2} \right) 
  \sum_s \frac{n_s}{C_{ns}} \exp\left[-\frac{k \cdot u_s}{T_s}\right] 
  \exp\left[-\frac{q \cdot u_s}{2 T_s} \right]\, \exp(iqx_s)  = \\
& 
      (k^\mu\, \hat\sigma_\mu) \ \left(2 \pi R^2 \right)^{3/2} 
\exp\left(-\frac{R^2}{2} q^2\right) \frac{n_s}{C_{ns}}
\exp\left[- \frac{k^0 \gamma_s}{T_s}\right]
\exp\left[- \frac{q^0}{2} \frac{\gamma_s}{T_s}\right]
\exp(i q^0 x_s^0) \times \\
& \left[
  \exp\left[ \frac{\vec k \vec u_s}{  T_s}  \right] 
  \exp\left[ \frac{\vec q \vec u_s}{2 T_s}  \right]
           \exp(-i \vec q \vec x_s) + 
  \exp\left[ -\frac{\vec k \vec u_s}{  T_s}  \right] 
  \exp\left[ -\frac{\vec q \vec u_s}{2 T_s}  \right]  
           \exp(i \vec q \vec x_s) \right],
\end{split}
\label{J2s3M} 
\ee
where the factor $\exp(i q^0 x_s^0)$ can be dropped if the FO time 
distribution is simultaneous for the two sources,
because then $x_s^0 = 0$.
Consequently, if the two sources have the same parameters, just
opposite locations with respect to the center, and opposite velocities,  
then the correlation function from Eq. (\ref{C-def}) is
\be
 C(k,q) =  1 +   \exp(-R^2 q^2) 
\frac{
 \cosh\left( \frac{2\vec k \vec u_s}{T_s} \right) +
 \cosh\left( \frac{ \vec q \vec u_s}{T_s} \right) 
  \cos( 2\vec q \vec x_s)  
}{  \cosh\left( \frac{2 \vec k \vec u_s}{T_s} \right) +1  } \ .
\label{C2ms}
\ee
If the velocity of the two sources vanishes this results returns 
the expression for two static sources obtained in Ref.~\refcite{Cso-5}.

\begin{figure}[ht] 
\begin{center}
      \includegraphics[width=7.6cm]{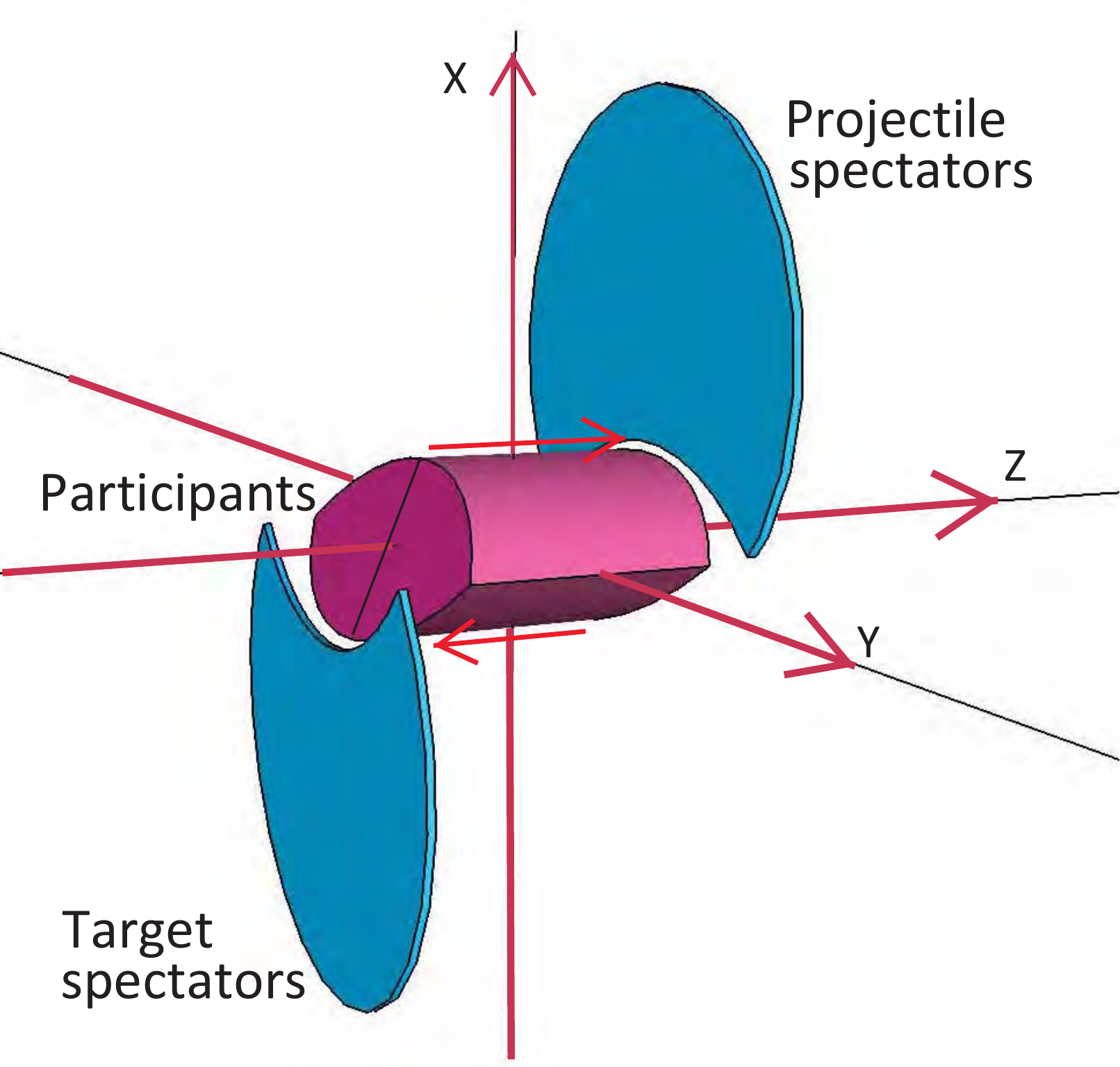}
\end{center}
\caption{
(Color online)
Typical orientation of the spatial axes in case of an 
ultra-relativistic heavy ion reaction shortly after the
impact. In the configuration space the projectile and
target appear to be flat due to the Lorentz contraction.}
\label{F-3}
\end{figure}

If we have two sources placed at $ x = \pm d_z$, 
and with the velocity in the $\pm x$-direction, $\pm v_x$, 
then the correlation function for the $k_x$ direction
becomes:
%
%
\begin{equation}
\begin{split}
C(k_x,q_x)  =\ & 1 + \exp(-R^2 q^2_x) \ 
\frac{
 \cosh\left( \frac{2 \gamma k_x  v_x}{T_s} \right) +
 \cosh\left( \frac{\gamma q_x  v_x}{T_s} \right)   
}{  \cosh\left( \frac{2 \gamma k_x  v_x}{T_s} \right) +1  } \ , \\
C(k_x,q_y)  =\ & 1 + \exp(-R^2 q^2_y)\ , \\
C(k_x,q_z)  =\ & 1 + \exp(-R^2 q^2_z) \ 
\frac{
 \cosh\left( \frac{2 \gamma k_x  v_x}{T_s} \right) +
 \cos\left( 2 q_z  d_z \right)   
}{  \cosh\left( \frac{2 \gamma k_x  v_x}{T_s} \right) +1  } \ .
\end{split}
\label{Oth-conf}
\end{equation}
%
Other equations for sources in different locations or the correlation 
function for $k_y$ and $k_z$ can be found using Eq. (\ref{C2ms}).
\begin{figure}[ht] 
\begin{center}
      \includegraphics[width=7.6cm]{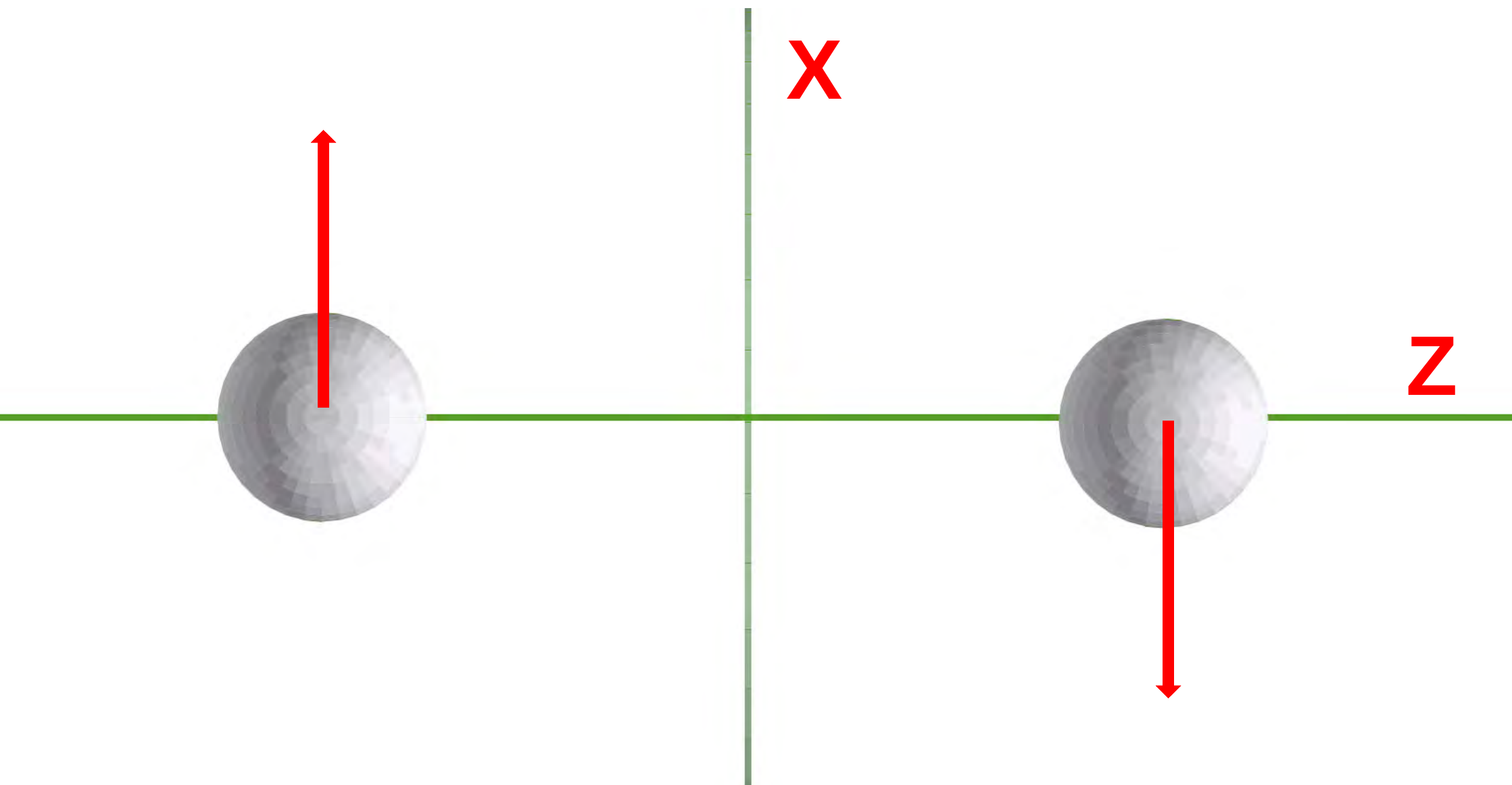}
\end{center}
\vskip -6mm
\caption{ (color online)
The correlation functions, $C(k,q)$ for two moving sources where the 
displacement of the sources is in the $z$-direction, and the 
center-of-mass momentum, $\vec k$, of emitted particles is in the 
 $x$-direction.
The {\it dashed green lines} are 
for the relative momentum, $q_z$, the {\it solid red line} is for $q_y$
and {\it dotted blue lines} are for $q_x$. 
For large values of the center-of-mass momentum $k_x$ 
the correlation functions $C(k_x,q_x)$ and $C(k_x,q_z)$ 
will approach the correlation
function $C(k_x,q_y)$ \,({\it red line}). 
For $q_x$ ({\it blue lines}) the displacements are  
$d_x = 1,0$ fm, 
and for $q_z$ ({\it green lines}) the velocity is chosen such that
$\gamma v_z /T_s = 1.0$ fm. The values of $k_z$ are for the 
{\it blue lines}: 0.25, 0.5, 0.75, 1.0, 1.25 and 2.0 fm$^{-1}$ and for the 
{\it green lines}: 0.25, 0.5, 0.75, 1.0 and 1.5 fm$^{-1}$.}
\label{F-2}
\vskip -3mm
\end{figure}
The correlation function for different source locations and velocities 
are similar. The cosine term appears in the same direction as the axis at 
which the sources are located and the hyperbolic cosine in the 
direction of the velocity.
The distribution does depend on the
magnitude of the flow velocity, $v_x$, but not on its direction!
This arises from the fact that the detectors are {\bf assumed to be}
reached from both sides of the system with opposite velocities
with equal probability.
Unfortunately the dominant direction of flow (see Fig. \ref{F-3}) 
is the beam direction ($z-$direction), where we have no possibility
to place high acceptance detectors. At the same time the strongest
effect of the flow appears in this direction.

In the case of Fig. \ref{F-2} the flow has
the most dominant effect in the $k_x$-direction, which is accessible
for detection. The $x$-directed flow, however, is more sensitively
dependent on secondary effects, like the Kelvin-Helmholtz Instability
\cite{hydro2}.

In this configuration of the sources the magnitude of the flow velocity
makes visible change in $C(k,q)$, in the $(k_x, q_x)$-direction also, 
which is
detectable by the usual detector configurations. Still the direction 
of the rotation does not appear in the observables, eq. (\ref{C2ms}), with the approach 
presented here. 

For these two-particle correlation measurements it is necessary to
identify independently, event by event the global collective 
reaction plane azimuth, $\Psi_{RP}$, experimentally and the
corresponding  event by event center of mass of the system
(e.g. with the method in Ref.~\refcite{Eyyubova-1,Eyyubova-2}). Knowing these we can
identify the $k_x$-direction (and the $k_y$-direction also).

It is important to mention that to detect rotation the accurate
identification of the reaction plane and its proper orientation 
is necessary. Furthermore,
not only the reaction plane with proper direction but also
the event by event center of mass (c.m.) should be identified
\cite{Eyyubova-1,Eyyubova-2}. This hardly ever done!  In both cases the use of
zero degree calorimeters provide an adequate tool as these
are sensitive to the spectator residues.

\subsection{Four Fluid Cell Sources}

\begin{figure}[ht] 
\begin{center}
      \includegraphics[width=7cm]{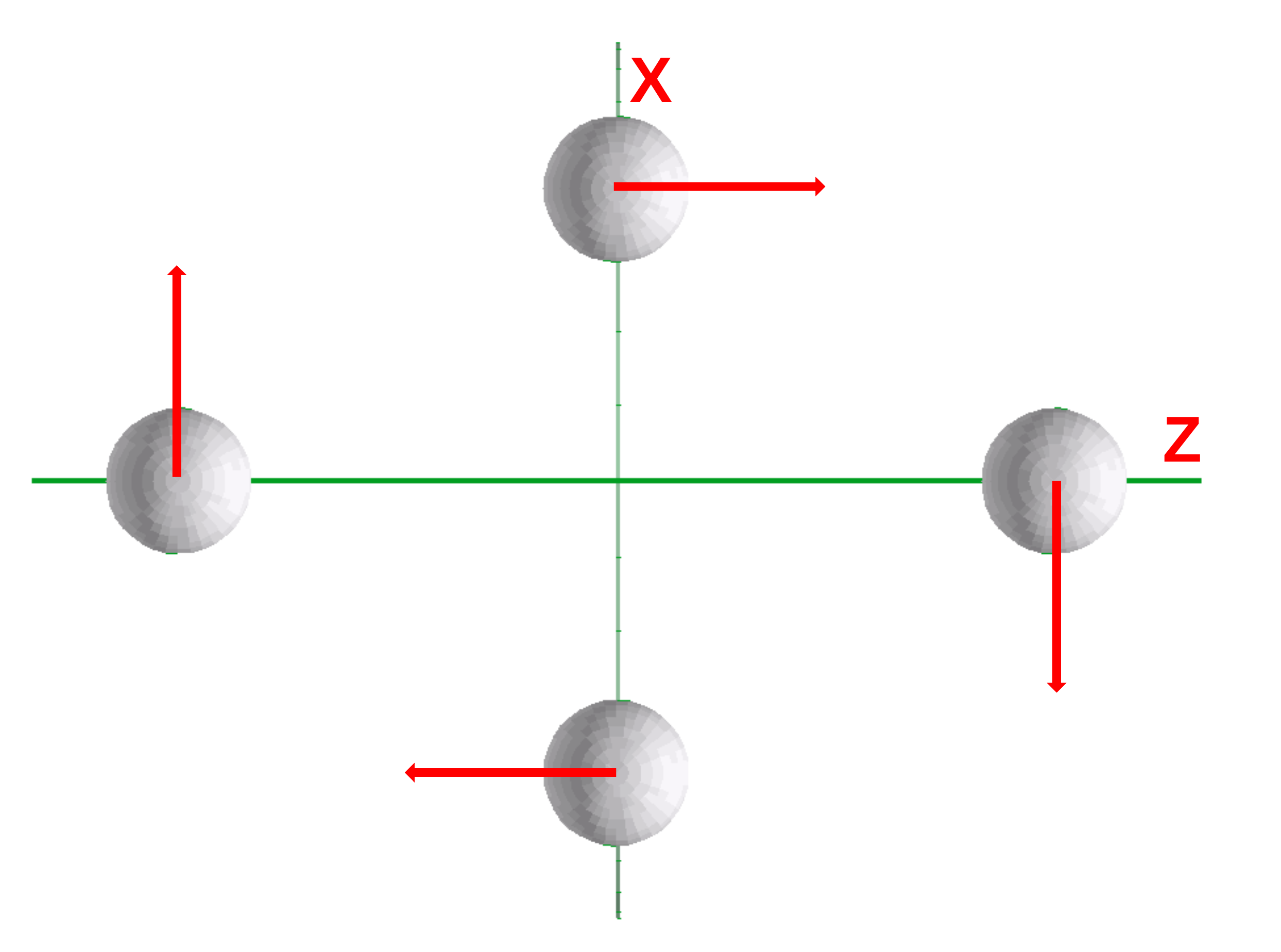}
\end{center}
\vskip -6mm
\caption{ (color online)
Four moving sources in the reaction ($[x-z]$) plane,
one pair, $s1$, is separated in the $x-$ directions and the other, $s2$,
is in the $z-$ direction. The sources are moving in the
directions indicated by the (red) arrows, $\pm \vec u_{s1}$ for the 1st pair
and $\pm \vec u_{s2}$ for the other.}
\label{F-4}
\end{figure}

Four sources can be treated as a combination
of two moving double source systems. We use the same parameters 
as under paragraph \ref{Ltsm}, where $s_1$ and $s_2$ will be the two 
different pairs of sources with different locations and velocities.
See Fig. \ref{F-4}.
 
So we obtain the correlation function for four sources as:
\be
\begin{split}
C(k,q) & =  exp(-R^2 q^2) \times  \\
& \left[
\cosh\left(\frac{2 \vec k \cdot \vec u_{s_1}}{T_s} \right)+ 
\cosh\left(\frac{\vec q \cdot \vec u_{s_1}}{T_s} \right)
\cos(2\vec q \cdot \vec x_{s_1})+ 
\right. \\
& \cosh\left(\frac{2 \vec k \cdot \vec u_{s_2}}{T_s} \right)
\cosh\left(\frac{\vec q \cdot \vec u_{s_2}}{T_s} \right)
\cos(2\vec q \cdot \vec x_{s_2}))  +  \\
& 2\cosh\left(\frac{\vec k \cdot (\vec u_{s_1} - \vec u_{s_2})}{T_s}\right)
\cosh\left(\frac{\vec q \cdot (\vec u_{s_1} + \vec u_{s_2})}{2 T_s}\right)
\cos(\vec q \cdot (\vec x_{s_1} + \vec x_{s_2}))  +  \\
& \left.
2\cosh\left(\frac{\vec k \cdot (\vec u_{s_1} + \vec u_{s_2})}{T_s}\right)
\cosh\left(\frac{\vec q \cdot (\vec u_{s_1} - \vec u_{s_2})}{2 T_s}\right)
\cos(\vec q \cdot (\vec x_{s_1} - \vec x_{s_2}))
\right] \times  \\
& \left[
\cosh\left(\frac{2 \vec k \cdot \vec u_{s_1}}{T_s} \right)+ 
\cosh\left(\frac{2 \vec k \cdot \vec u_{s_2}}{T_s} \right) +
2\cosh\left(\frac{ \vec k \cdot (\vec u_{s_1} + \vec u_{s_2})}{T_s} \right) +
\right. \\
& \left.
2\cosh\left(\frac{ \vec k \cdot (\vec u_{s_1} - \vec u_{s_2})}{T_s} \right) +
2  \right]^{-1} \ .
\end{split}
\label{C4ms1}
\ee
If $s_1 = s_2$ then we recover Eq. (\ref{C2ms}).

In the case of a rotating but symmetric system the displacements
and velocities are of equal magnitude and are 
orthogonal to each other in the two pairs:
$\vec x_{s_1} \perp \vec x_{s_2}$ and
$\vec u_{s_1} \perp \vec u_{s_2}$. Thus a simple sign change of the 
velocity for one of the pairs or both does not change the result,
and so the rotation can be identified, 
but this evaluation does
not provide sensitivity to the direction of the rotation.
The reason is in the over-simplified freeze out assumption as we mentioned
already at the end of paragraph \ref{Ltsm}.

{\bf Four Sources with Flow Circulation:}
Recent fluid dynamical studies indicate
\cite{hydro1,hydro2}, that due to the initial
shear and angular momentum the early fluid dynamical development
has significant flow vorticity and circulation on the reaction
plane. These were recently evaluated \cite{CMW12}. At the present 
LHC Pb+Pb collision energy in the mentioned fluid dynamical model
calculation the maximum value of vorticity, $\omega$, was found 
exceeding $3$ c/fm , and the circulation after $6$ fm/c flow
development and expansion was still around 4-5 fm$\cdot$c. 
This vorticity in the reaction plane was more than an order of 
magnitude bigger than in the transverse
plane estimated from random fluctuations \cite{FW11-1,FW11-2}.
\begin{figure}[ht] 
\begin{center}
      \includegraphics[width=7.6cm]{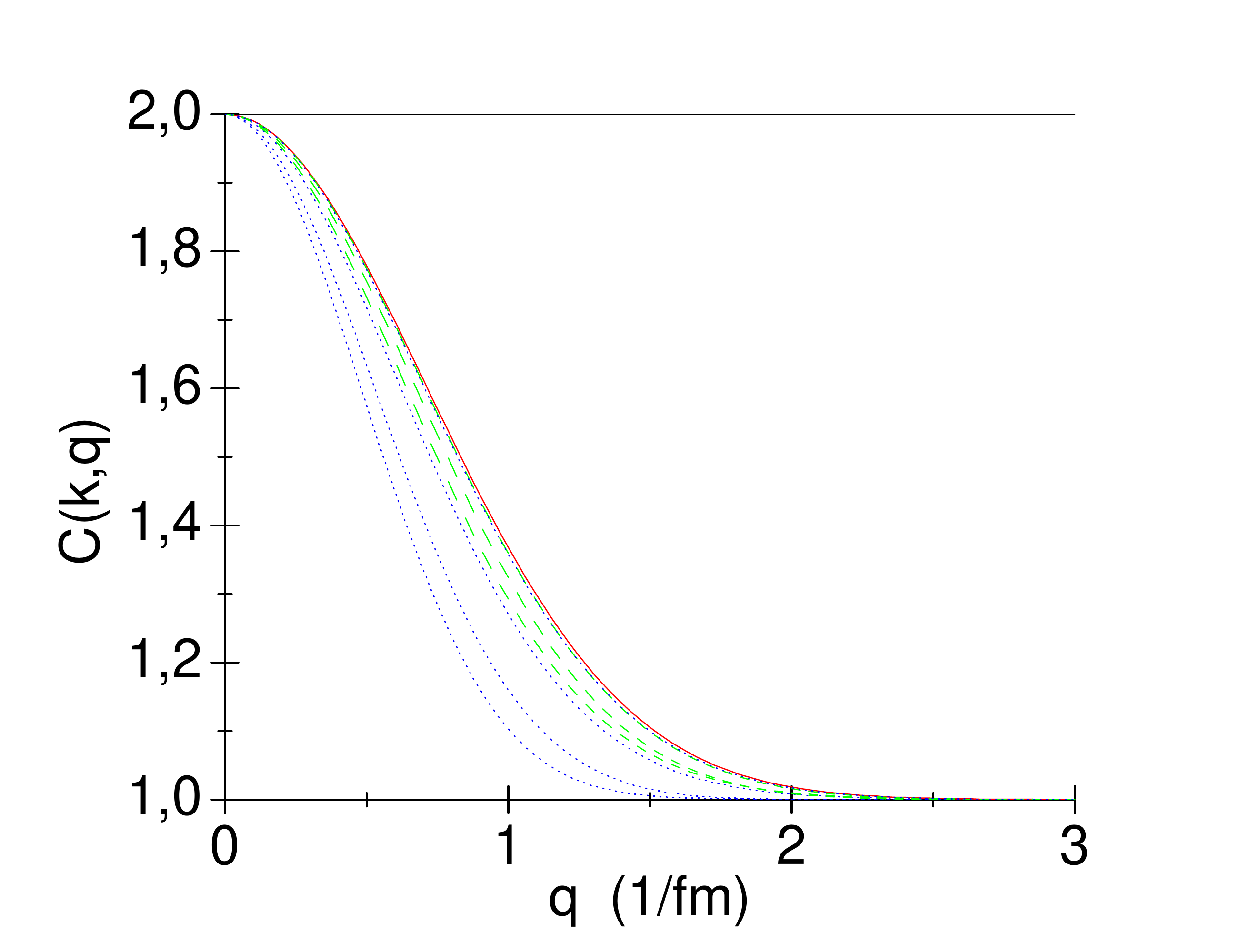}
\end{center}
\vskip -6mm
\caption{ (color online)
The correlation functions, $C(k,q)$, for 4 sources 
where the displacement is such that there is one 
source pair on the x-axis and one the z-axis, 
the center-of-mass momentum of the emitted particles, 
$\vec k$, is in the x-direction. 
The {\it dotted blue lines} are for the velocity $v=0.5$\,c 
and displacement $d_x = d_z = 1.6$ fm. 
The {\it dashed green lines} are for the velocity $v=0.8$\,c 
and displacement $d_x = d_z = 1.0$ fm. $T_s = 0.20$ GeV.
In both cases the circulation is $\Gamma = 5$ fm$\cdot$c.
The values of $k_x$ are for the {\it blue lines}: 
0.5, 1.5, 3.0 and 6.0 fm$^{-1}$ and for the {\it green lines}: 
0.5, 1.5 and 3.0 fm$^{-1}$.
The {\it solid red line} is the correlation function $C(k_x,q_y)$. 
For large values of the center-of-mass momentum $k_x$ the 
correlation functions $C(k_x,q_x)$ and $C(k_x,q_z)$ 
will approach the correlation function $C(k_x,q_y)$. 
The larger displacement and smaller rotation velocity
leads to stronger deviation from the unaffected
correlation function $C(k_x,q_y)$.  }
\label{F-5}
\end{figure}

In this section we will look at the four source 
correlation function with similar circulation as in the above
mentioned fluid dynamical model estimates in the reaction plane \cite{CVW2014}. 
See Fig. \ref{F-4}.
We will simulate a circulation value $\Gamma = 5 fm \cdot c$.
We use Eq. (\ref{C4ms1}) where the 
center-of-mass momentum, $\vec k$ points in the $x-direction$.
Since the position and velocity are of the same 
value and because of symmetry the correlation functions 
$C(k_x,q_x)$ and $C(k_x,q_z)$ provide the same values. 
So we take the correlation function $C(k_x,q_x)$ and 
we have afterwards some simplifications. See Fig. \ref{F-5}.
\be
\begin{split}
& C(k_x,q_x) = 1 + \exp(-R^2 q^2) 
\left[ 1+\cos(2 q_x d) + \cosh\left(2\frac{k_x \gamma v_x}{T_s}\right) + 
\cosh\left(\frac{q_x \gamma v_x}{T_s}\right) + \right. \\
&
\left. 4 \cosh\left(\frac{k_x \gamma v_x}{T_s}\right) 
\cosh\left(\frac{q_x \gamma v_x}{2 T_s}\right) \cos(q_x d) \right]  
\left[\cosh\left(\frac{2 k_x \gamma v_x}{T_s}\right) + 
4 \cosh\left(\frac{k_x \gamma v_x}{T_s}\right) + 3 \right]^{-1}
\end{split}
\ee
\begin{figure}[ht] 
\begin{center}
      \includegraphics[width=7.6cm]{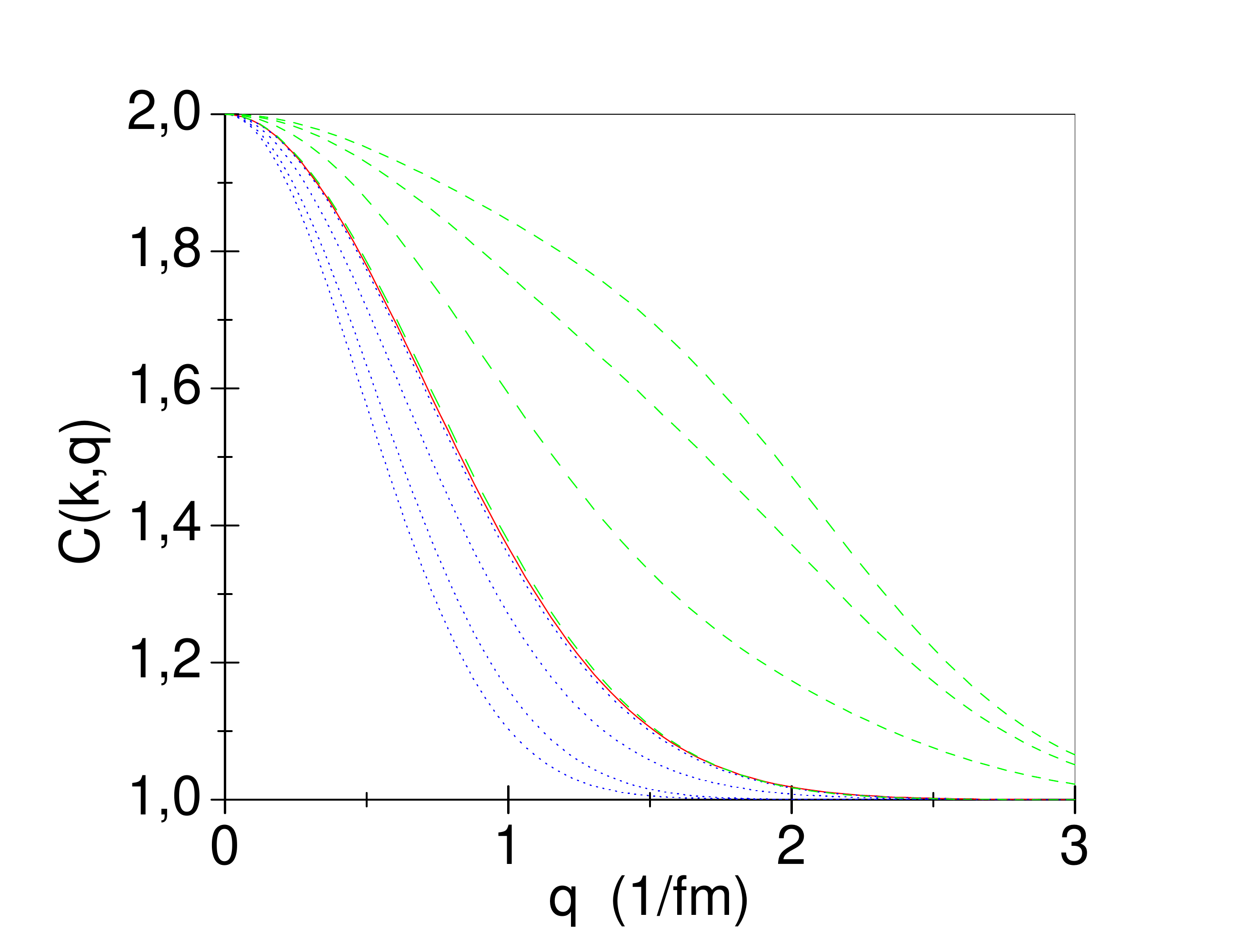}
\end{center}
\vskip -6mm
\caption{ (color online)
The correlation functions, $C(k,q)$, for 4 sources 
where the displacement is such that there is one 
source pair on the x-axis and one the z-axis, 
the center-of-mass momentum of the emitted particles, 
$\vec k$, is in the x-direction. 
The {\it dotted blue lines} are for the velocity $v=0.5$\,c 
and displacement $d_x = d_z = 1.6$ fm (same as in the previous figure). 
The {\it dashed green lines} are for the velocity $v=0.95$\,c 
and displacement $d_x = d_z = 0.84$ fm, $T_s = 0.20$ GeV.
The values of $k_x$ are: for the {\it blue lines} 
0.5, 1.5, 3.0 and 6.0 fm$^{-1}$ and for the {\it green lines} 
0.1, 0.25, 0.5 and 1.5 fm$^{-1}$.
The {\it solid red line} is the correlation function $C(k_x,q_y)$. 
For large values of the center-of-mass momentum $k_x$ the 
correlation functions $C(k_x,q_x)$ and $C(k_x,q_z)$ 
will approach the correlation function $C(k_x,q_y)$. 
Now for the dashed green lines with even higher velocity and 
smaller displacement, the deviation is significant and it is in the 
positive direction.
}
\label{F-6}
\end{figure}
For $C(k_x,q_y)$ we have the same result as 
we had for the two moving sources. Here the flow and displacement have no
effect.  See Fig. \ref{F-6}.

Let us look at comparisons for similar circulations and 
for similar displacements. 
We see that an 
increase in the displacement of the sources gives a increase in 
the apparent size of the system (narrower $q-$distribution). 
We also see that the measured size of the system increases with 
decreasing velocity.
At the same time the shape of correlation functions are becoming 
less and less Gaussian as the flow velocities increase. At the same time
the structure of the correlation function is also very different in 
different directions, which is not the case for spherical 
or linear expansion. This indicates that the rotating 
system contributes to essential non-Gaussian modifications,
which can be seen directly in the correlation function, but they would
become invisible if we would like to fit these data with a set of
Gaussians. Earlier works studied the correlation function at different 
angles or pseudorapidities with Gaussian parametrizations
\cite{DM95,Sin89},
however, for rotating systems this is not the most sensitive way of
presenting the results.

\section{Asymmetric Sources}
\label{ASm}
\vskip -3mm

We have seen in the previous few source model examples that
a highly symmetric source may result in 
correlation functions that are sensitive to rotation, however, 
these results were not sensitive 
to the direction of the rotation, which seems to be unrealistic.
We saw that this result is a consequence of the assumption that
both of the members of a symmetric pair contribute equally to the
correlation function even if one is at the side of the system
facing the detector and the other is on the opposite side. 
The expansion velocities are also opposite at the opposite sides. The
dense and hot nuclear matter or the Quark-gluon Plasma are
strongly interacting, and for the most of the observed particle types
the detection of a particle from the side of the system, -- which
is not facing the detector but points to the opposite direction, --
is significantly less probable. The reason is partly in the diverging 
velocities during the expansion and partly to the lower emission
probability from earlier (deeper) layers of the source from the
external edge of the timelike (or spacelike) FO layer. 
This feature is recognized for
a long time and discussed in detail by now. This topic has an extended
literature, and this feature destructs the symmetry of emission
of from source pairs at the opposite sides of the system
\cite{Sin89, M-2, M-3, M-4, CF, Si89-1,Si89-2,Si89-3, Bugaev, ALM99, ACG99, 
MAC99, Cs02, MAA03, TC04, MCM05, MCM06b}.

For the study of realistic systems where the emission is dominated
by the side of the system, which is facing the detector, 
we cannot use the assumption of the symmetry among pairs or 
groups of the sources from opposite sides of the system. Even if the
FO layer has a time-like normal direction, $\hat\sigma^\mu$ the 
$(k^\mu \hat\sigma^\mu)$ factor yields a substantial emission
difference between the opposite sides of the system. 
This simple freeze-out emission probability factor was later taken into 
account in earlier theoretical two particle 
correlation studies, see e.g. Ref.~\refcite{CVW2014} and Ref.~\refcite{Cso-5}. 

\subsection{The Emission Probability}\label{TEP}

It was first recognized that the freeze out with the
Cooper-Fry description \cite{CF}, may lead to negative
contributions for particles, which move towards the center of 
the system and not in the direction out, towards the detectors.
The first proposal to remedy this problem came from Bugaev
\cite{Bugaev}, which led to the introduction of an improved
post freeze out distribution in the Cooper-Frye description,
first with the Cut-J\"uttner distribution \cite{Bugaev,ACG99}
and then by the Cancelling-J\"uttner distribution \cite{TC04}.

Later the necessity to introduce
an escape probability, $P_{esc}$ was pointed out. 
The escape probability was then introduced and analysed
in a series of publications \cite{M-2,M-3,M-4,MCM06b},
in transport theoretical approaches. It was pointed out that
even if the pre FO distribution is a locally equilibrated
isotropic distribution, the freeze out process and the
escape probability will provide a nonisotropic distribution.

The escape probability introduced in the works 
\cite{M-2,M-3,M-4,MCM06b}, for a space-time surface layer 
of the system of thickness $L$, pointing in the four
direction $\hat\sigma^\mu$ was given at a point $x^\mu$ inside the
freeze out layer as
\be
P_{esc}(x) \ \propto \ 
\left(\frac{L}{L-x^\mu \hat\sigma_\mu}\right)
\left(\frac{p^\mu \hat\sigma_\mu}{ p^\mu u_\mu}\right)
\Theta(p^\mu \hat\sigma_\mu) \,,
\ee
where $p^\mu$ is the 
momentum of the escaping particle, $u^\mu(x)$ is the
local flow velocity and $s=x^\mu \hat\sigma_\mu$ is the distance of the
emission point from the inside boundary of the layer. The first
multiplicative term describes higher emission probability
to the particles, which are emitted closer to the outside boundary
of the layer, the second multiplicative term describes the
higher emission probability for the particles, which move in the
normal direction of the surface, because these should cross
less material in the layer. The last term secures that only those 
particles can escape, which move outwards through the layer.

The correlation function, $C(k,q)$ is always measured
in a given direction of the detector, $\vec k$. Obviously
only those particles can reach the detector, which 
satisfy   $ \ k^\mu \hat\sigma_\mu \ > \ 0 $. Thus 
in the calculation of  $C(k,q)$ for a given $\hat{\vec k}$-
direction we can exclude the parts of the freeze out
layer where $ \ k^\mu \hat\sigma_\mu \ < \ 0 $ (see Eq. (10)
of Ref.~\refcite{Sin89} or Ref.~\refcite{Bugaev}. For time-like 
FO a simplest approximation for the emission possibility is
$
P_{esc}(x) \ \propto \  k^\mu u_\mu(x) 
$
\cite{Cso-5}. 

%

\subsection{Emission probabilities for few sources}\label{EPFS}

{\bf Two sources:}

In the configuration where two sources are in the
beam-, $z-$direction, the observation can be in
different $\hat{\vec k}$-directions. If $\hat{\vec k}$ points
into the $\pm y-$direction, then the probabilities must
be identical so emission probabilities do not lead to any change.

If $\hat{\vec k}$ points into the $\pm x-$direction,
then one of the sources is closer to the detector and may
shadow the more distant one. Thus, we can just introduce two
positive weight factors so that $w_c$ is the weight for the cells closer
to the detector and $w_s$ is for the cells which are far from
the detector measuring the average momentum $\vec k$. These
weights are the same for the calculation of the nominator and denominator
of the correlation function, so their normalization does not
influence the correlation function. 

As not all emitted
particles reach a given detector the normalization is also dependent
on the direction of the detector. Thus, we evaluate the 
correlation function this way. This immediately changes the earlier
result (\ref{Oth-conf}), because it breaks the symmetry between the
two sources. 
\begin{figure}[ht] 
\begin{center}
      \includegraphics[width=6cm]{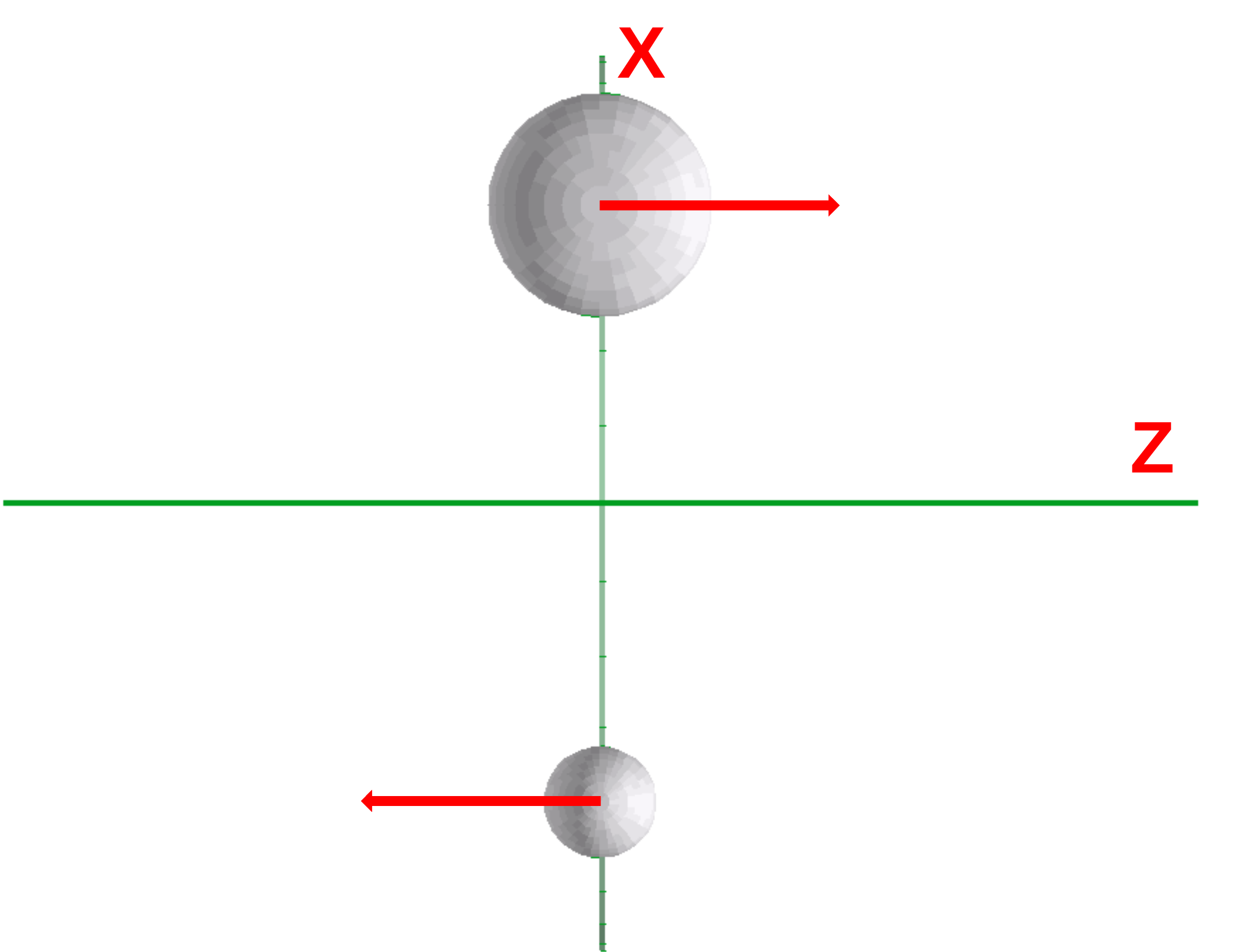}
\end{center}
\vskip -6mm
\caption{ (color online)
Two moving sources in the reaction ($[x-z]$) plane,
separated in the $x-$ direction (case (ii) in the text). 
The sources are moving in the
directions indicated by the (red) arrows. The detector is in the 
positive $x-$direction, thus the source on this side has more
dominant emission into this direction, and this is indicated 
by the bigger size of the source on this side.}
\label{F-7}
\end{figure}
We can simply repeat the calculation
for two moving sources in section \ref{Ltsm}, modifying the derivation 
of Eq. (\ref{C2ms}) and obtain the general result 
\be
\begin{split}
& \ \ \ \ \ \ \left. C(k,q)\!\!\phantom{\frac{1}{2}}\!\!\right|_{+x} 
= 1 + \exp(-R^2 q^2) \ \times\\ 
& 
\frac{
 w_c^2 e^{ \frac{2\vec k \vec u_s}{T_s}}  +
 w_s^2 e^{-\frac{2\vec k \vec u_s}{T_s}}  +
 2 w_c w_s \cosh\left( \frac{ \vec q \vec u_s}{T_s} \right) 
  \cos( 2\vec q \vec x_s)  
}{  
 w_c^2 e^{ \frac{2\vec k \vec u_s}{T_s}}  +
 w_s^2 e^{-\frac{2\vec k \vec u_s}{T_s}}  + 2 w_c w_s  } \ .
\end{split}
\label{C2msAs}
\ee
Note that this result is valid for the case when $\hat{\vec k}$ 
points to the $+x$ direction, because the weights depend on this
and $w_c > w_s$. See Fig. \ref{F-7}. The fact that the emission
from the source, which is closer to the detector is stronger
makes the direction of the flow detectable.

If we introduce the notation $w_c = 1+\epsilon$
and $w_s = 1-\epsilon$, the deviation from the symmetric
result will become apparent

\be
\begin{split}
\left. C(k,q)\!\!\phantom{\frac{1}{2}}\!\!\right|_{+x} 
& = 1 + \exp(-R^2 q^2) \times \\
& \frac{
 (1+\epsilon^2)  \cosh\left(\frac{2\vec k \vec u_s}{T_s}\right)  +
 2 \epsilon      \sinh\left(\frac{2\vec k \vec u_s}{T_s}\right)  +
 (1-\epsilon^2)  \cosh\left( \frac{ \vec q \vec u_s}{T_s} \right) 
  \cos( 2\vec q \vec x_s)  
}{  
 (1+\epsilon^2)  \cosh\left(\frac{2\vec k \vec u_s}{T_s}\right)  +
 2 \epsilon      \sinh\left(\frac{2\vec k \vec u_s}{T_s}\right)  +
 (1-\epsilon^2)  } \ .
\label{C2msAsw}
\end{split}
\ee
If $\epsilon \rightarrow 0$, i.e. if $w_c=w_s$, we recover the earlier
result, Eq. (\ref{C2ms}).
\begin{figure}[ht] 
\begin{center}
      \includegraphics[width=6cm]{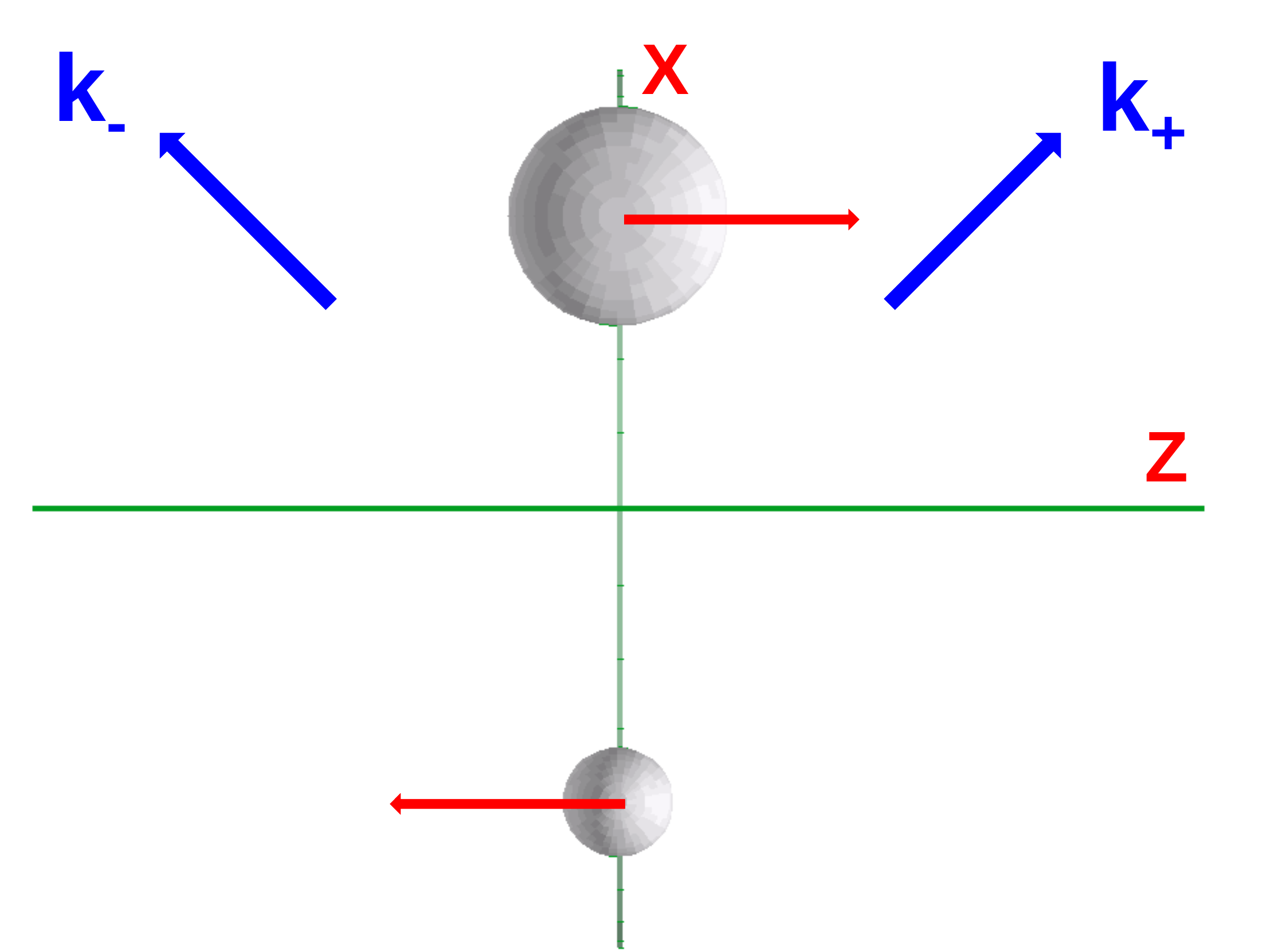}
\end{center}
\vskip -4mm
\caption{ (color online)
Two moving sources in the reaction ($[x-z]$) plane,
separated in the $x-$ direction. 
The sources are moving in the
directions indicated by the (red) arrows. The two "tilted" detector
directions are indicated by the (blue) arrows labeled with
$k_+$ and $k_-$.}
\label{F-8}
\end{figure}
If $\epsilon = 0$ we have the symmetric situation where both
sources have equal contribution, the asymmetric terms vanish, and
the result becomes to be symmetric for the change of the direction
of the flow velocity.
This result has terms, which change sign if the flow velocity, ${\vec u}_s$
changes sign. The result is valid only if the detector is in the 
$\hat{\vec k} = (1,0,0)$ direction. For this direction, however, if the
flow velocity points in the $z$-direction, i.e. orthogonal to $\vec k$
the asymmetric term does not provide any contribution, so
it will not show up in $C(k_x,\vec q)$. To circumvent this problem
we should study detector directions, which do not coincide with
the primary axes of the given event (where $x$ is the direction of 
the impact parameter vector, $\vec b$, pointing to the projectile;
$y$ is the other transverse direction; and $z$ is the direction of the 
projectile beam).

{\bf Correlation in Tilted Directions:}
The form of the correlation function is the same if $\vec k$ is in the 
same plane, the reaction plane, 
but it has a $z$ component also, i.e. $\vec k = (k_x, 0, \pm k_z)$.
This is possible for all LHC heavy ion experiments, ATLAS, CMS and
even ALICE, where the longitudinal acceptance
range of the TPC ($\Delta \eta \ < 0.8$) is the smallest. 
See Fig. \ref{F-8}.

Depending on the detector 
acceptance we should chose a detector direction where $|k_z|$ is as big
as the detector acceptance allows it.    
For this configuration the {\bf form} of the correlation function is 
the same as (\ref{C2msAsw})
\be
\left. C(k,q)\!\!\phantom{\frac{1}{2}}\!\!\right|_{+x,\pm z} =
\left. C(k,q)\!\!\phantom{\frac{1}{2}}\!\!\right|_{+x} \ ,
\ee
with keeping the different weights, $w_c, w_s$ or $\epsilon$ so that
the forward shifted and backward shifted directions have the same weights. 
These weights are not specified up to now.

For detection of the correlation function we have to introduce here
the usual, $\vec k$-dependent coordinate system to classify the
direction of $\vec q$.
Thus if 
\be 
\hat{\vec k}_\pm = (a,0,\pm b){\rm fm}^{-1} , \ 
k_x = a | \vec k | , \
k_z = \pm b | \vec k | , 
\label{kcomp}
\ee
where $a^2 + b^2 = 1$, see Fig. \ref{F-8},
then the difference vector, $\vec q$, can be measured in the directions
\be
\begin{split}
& \hat{\vec q}_{out}  = (a,0, \pm b) , \
q_x = a | \vec q | , \ q_z = \pm b | \vec q | \\
& \hat{\vec q}_{side} = (0,1,0) , \ \ \ q_y = | \vec q | \\
& \hat{\vec q}_{long} = (\mp b,0,a) , \ \ 
q_x = \mp b | \vec q | , \ q_z = a | \vec q | .
\end{split}
\label{qcomp}
\ee
This leads to the following correlation functions

\begin{equation}
\begin{split}
& C(k_{(\pm)},q_{out})  = 1 +  \exp(-R^2 q^2)\, \times \\
& \frac{
 (1{+}\epsilon^2) \cosh\left( \frac{2 \gamma k_z  v_z}{T_s} \right) +
 2 \epsilon       \sinh\left( \frac{2 \gamma k_z  v_z}{T_s} \right) +
 (1{-}\epsilon^2) \cosh\left( \frac{  \gamma q_z  v_z}{T_s} \right) 
                   \cos\left( q_x  d_x \right)   
}{  
 (1{+}\epsilon^2) \cosh\left( \frac{2 \gamma k_z  v_z}{T_s} \right) +
 2 \epsilon       \sinh\left( \frac{2 \gamma k_z  v_z}{T_s} \right) + 
  (1{-}\epsilon^2) } \ , \\
& C(k_{(\pm)},q_{side}) = 1 + \exp(-R^2 q^2)\ , \\
& C(k_{(\pm)},q_{long}) = 1 + \exp(-R^2 q^2)\, \times \\
&\frac{
 (1{+}\epsilon^2) \cosh\left( \frac{2 \gamma k_z  v_z}{T_s} \right) +
 2 \epsilon       \sinh\left( \frac{2 \gamma k_z  v_z}{T_s} \right) +
 (1{-}\epsilon^2) \cosh\left( \frac{  \gamma q_z  v_z}{T_s} \right) 
                   \cos\left( q_x  d_x \right)   
}{  
 (1{+}\epsilon^2) \cosh\left( \frac{2 \gamma k_z  v_z}{T_s} \right) +
 2 \epsilon       \sinh\left( \frac{2 \gamma k_z  v_z}{T_s} \right) + 
  (1{-}\epsilon^2) } \ .
\end{split}
\label{C2kx1}
\end{equation}
\begin{figure}[ht] 
\begin{center}
      \includegraphics[width=7.9cm]{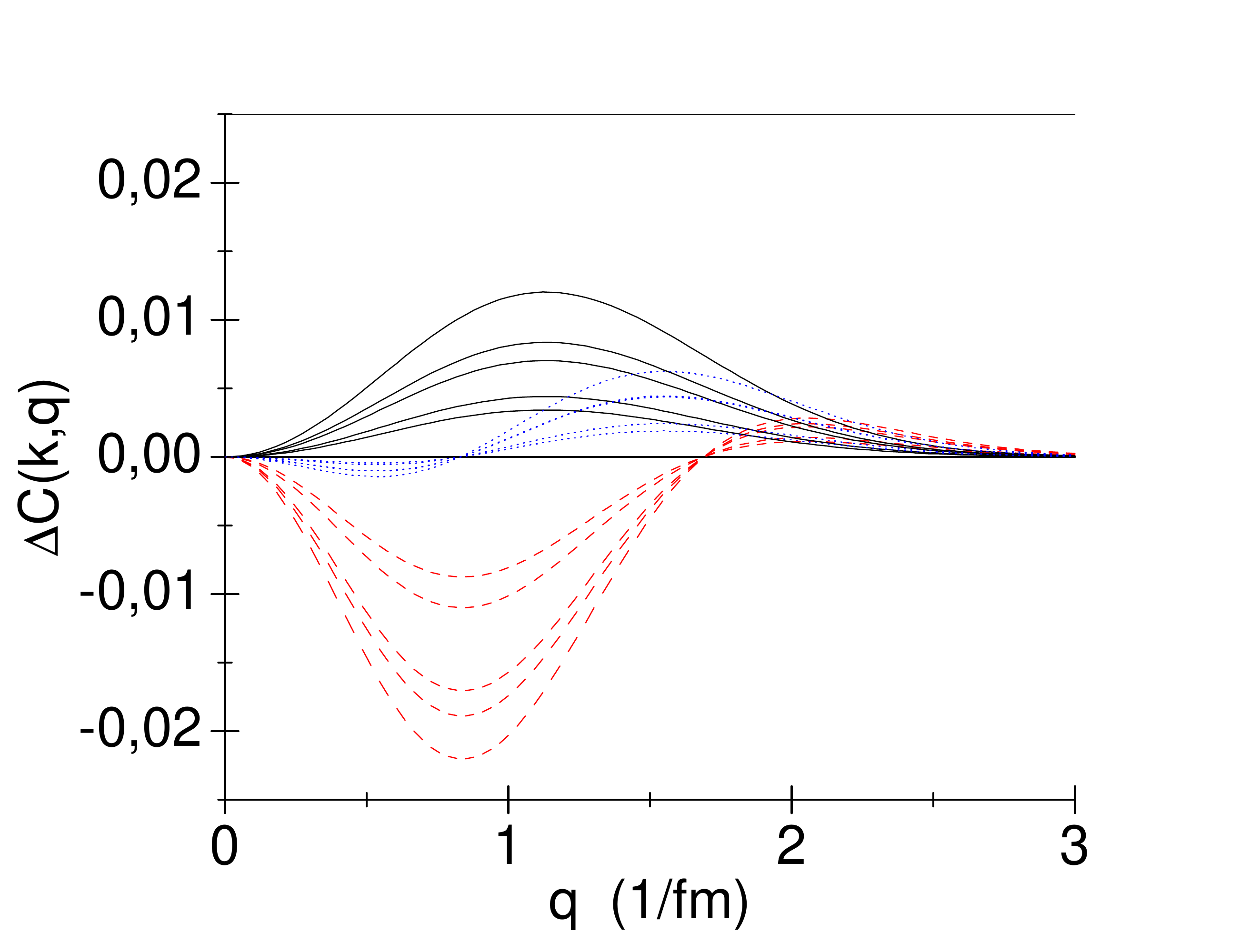}
\end{center}
\vskip -0.7cm
\caption{ (color online)
Difference of the forward and backward shifted 
correlation function,  $\Delta C(k_\pm,q_{out})$, 
for the value $\epsilon = 0.50$.
The {\it solid black lines} are for the velocity $v_z=0.5$\,c,
{\it dotted blue lines} are for the velocity $v_z=0.6$\,c and
{\it dashed red lines} are for the velocity $v_z=0.7$\,c.
Displacement is $d_x = 1.0$ fm, $T_s = 0.139$ GeV 
and $a = b = 1/\sqrt{2}$.
The values of $k$ are: for the {\it solid black lines} 
0.25, 0.50, 2.00, 2.75 and 3.50   fm$^{-1}$,
the {\it dotted blue lines}
0.25, 0.50, 1.00, 1.75 and 2.50 fm$^{-1}$,
and the {\it dashed red lines}
0.25, 0.50, 0.75, 1.25 and 1.75 fm$^{-1}$.}
\label{F-9}
\end{figure}

Although, it seems that  $C(k_{(\pm)},q_{out})$ and $C(k_{(\pm)},q_{long})$
are the same, this is in fact not the case, because the values of the
components of the different types of $\vec k$ and $\vec q$ are not the same
as described in Eqs. (\ref{kcomp},\ref{qcomp}). In all cases,
the out-, side- and long- $ q = | \vec q |$. We will also use the notation
$k = |\vec k|$ and $\gamma v_x = u_x$, $\gamma v_y = u_y$, $\gamma v_z = u_z$,
so that $\vec u_s = (u_x, u_y, u_z)$.
For example for the {\it out}
component the difference of the forward and backward shifted correlation 
functions is
\begin{equation}
\begin{split}
& 
\Delta C (k_\pm, q_{out}) \equiv   
C(k_{+},q_{out}) -  C(k_{-},q_{out}) = \\
& 
\frac{4 \exp(-R^2 q^2)\ 
  \epsilon  \sinh\left( \frac{2 u_z\, b k}{T_s} \right)\ 
  (1{-}\epsilon^2)\left[ 1 - 
   \cosh\left( \frac{ u_z\, b q}{T_s} \right) 
                   \cos\left(a q d_x \right) \right]
}{
 \left[(1{+}\epsilon^2) \cosh\left(\frac{2 u_z\, b k}{T_s}\right) 
      +(1{-}\epsilon^2) \right]^2 - 
  4 \epsilon^2  \sinh^2\left(\frac{2 u_z\, b k}{T_s} \right)}.
\end{split}
\label{DCpm}
\end{equation}
As Eq. (\ref{DCpm}) and Fig. \ref{F-9} show, the Differential Correlation 
Function (DCF), 
\linebreak 
$\Delta C(k_\pm,q_{out})$, 
is sensitive to the speed and direction 
of the rotation, and it is also sensitive to the amount of the 
tilt in the directions of the detection, regulated here by the
parameters $a$ and $b$.
$\Delta C(k_\pm,q_{out})$ tends to zero both if $q \rightarrow 0$
and if $q \rightarrow \infty$. 
\begin{figure}[ht] 
\begin{center}
      \includegraphics[width=6cm]{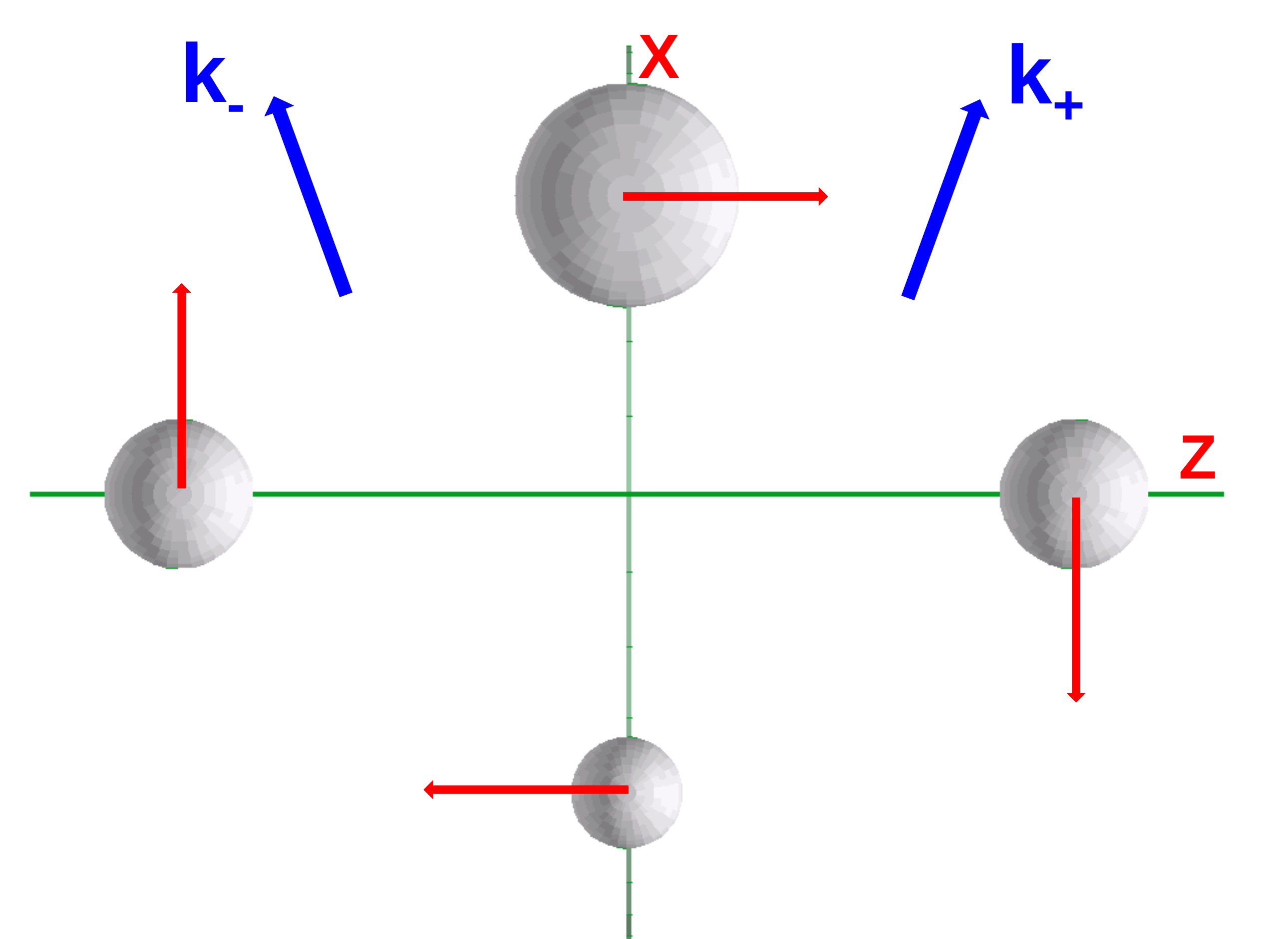}
\end{center}
\vskip -4mm
\caption{ (color online)
Four moving sources in the reaction ($[x-z]$) plane,
separated in the $x-$ and $z-$ directions. 
The sources are moving in the
directions indicated by the (red) arrows. The "tilted" detector
directions are indicated by the (blue) arrows.}
\label{F-10}
\end{figure}
In case if the detector has a narrow pseudorapidity acceptance, then
$\vec k_\pm$ is close to $k_x$, i.e. $b \ll a$ and then the weights 
are maximal for the source in the $x-$direction.

The $\sinh( 2 u_z\, b k /T_s )$ term in eq. (\ref{DCpm}) changes sign in the 
nominator when $u_z$ changes sign the difference of the two 
correlation functions, $\Delta C(k_\pm,q_{out})$ changes sign also
because all other terms are symmetric to the sign change of the 
velocity. 

This is an important observation as we can detect the
direction and magnitude of the rotation in the reaction plane.
This difference is also increasing with the longitudinal shift, $b$, 
of the average momentum vector, $\vec k$, so that detectors with
larger pseudorapidity acceptance can detect the rotation better. 

  In order to perform this measurement, one has to determine the
global reaction plane (e.g. from spectator residues in the ZDCs), and 
determine the projectile side of this plane as it was mentioned earlier. 
Furthermore
the event by event center of mass should also be identified
(using e.g. the method shown in Ref.~\refcite{Eyyubova-1,Eyyubova-2}). This will be the 
positive $x$-direction. Then the correlation function can be measured
for four different $\vec k$-directions in the global reaction plane.
These four directions are shifted forward and backward from the
center of mass symmetrically on the projectile side, and there should
be a symmetric pair of detection points in the target side of the
reaction plane too. In the realistic case with taking into account 
asymmetries that  arise from the freeze out, the proper determination 
of the c.m. and reaction plane are even more important.

The $\vec k$ directions opposite to each other across the c.m. point
give the same result, while the difference, $\Delta C(k_\pm,q_{out})$, 
between the Forward (F) and 
Backward (B) shifted contributions will characterize the speed and direction
of the rotation. This symmetry can be used to eliminate the contribution
from eventual random fluctuations.  The observed F/B asymmetry depends on
the parameters $\epsilon$, $v_z$ and $d_x$, these can be estimated by
measuring the correlation functions at all possible moments $\vec k$.

\subsection{Emission from four sources}\label{Ef4S}

With four sources we can illustrate the possibilities of 
differential HBT method studies in different directions.
The correlation functions can be calculated in general 
for four sources and two detector positions.  This can then
be applied to different detector configurations.

The out component of the four source correlation function with 
weight factors $\omega_a$, $\omega_b$, $\omega_c$, $\omega_d$ 
can be found by using the same method as for the two source case.
\begin{figure*}[h] 
\begin{minipage}{.44\textwidth}
\begin{center}
      \includegraphics[width=6cm]{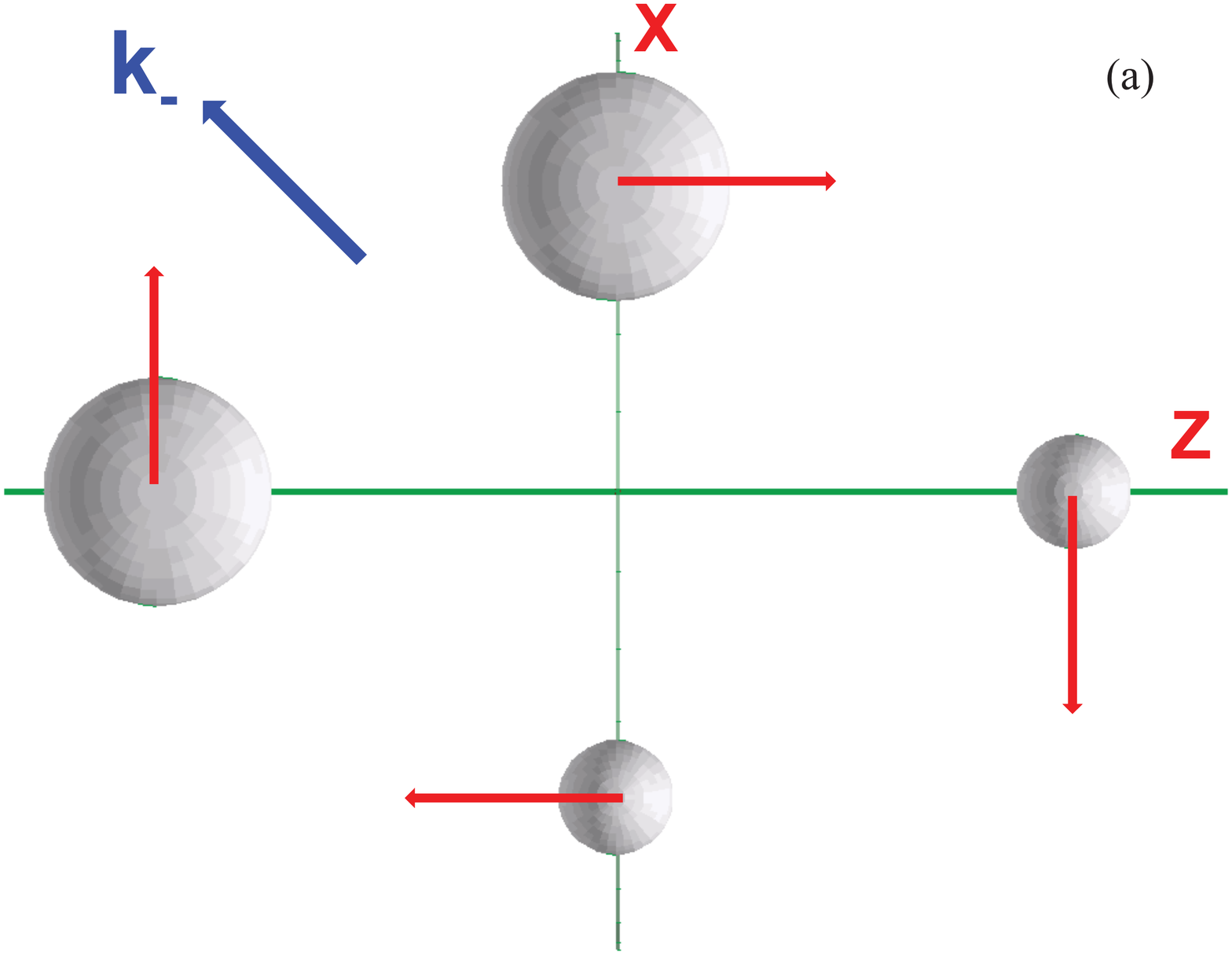}
\end{center}
\end{minipage}
\begin{minipage}{.44\textwidth}
\begin{center}
      \includegraphics[width=6cm]{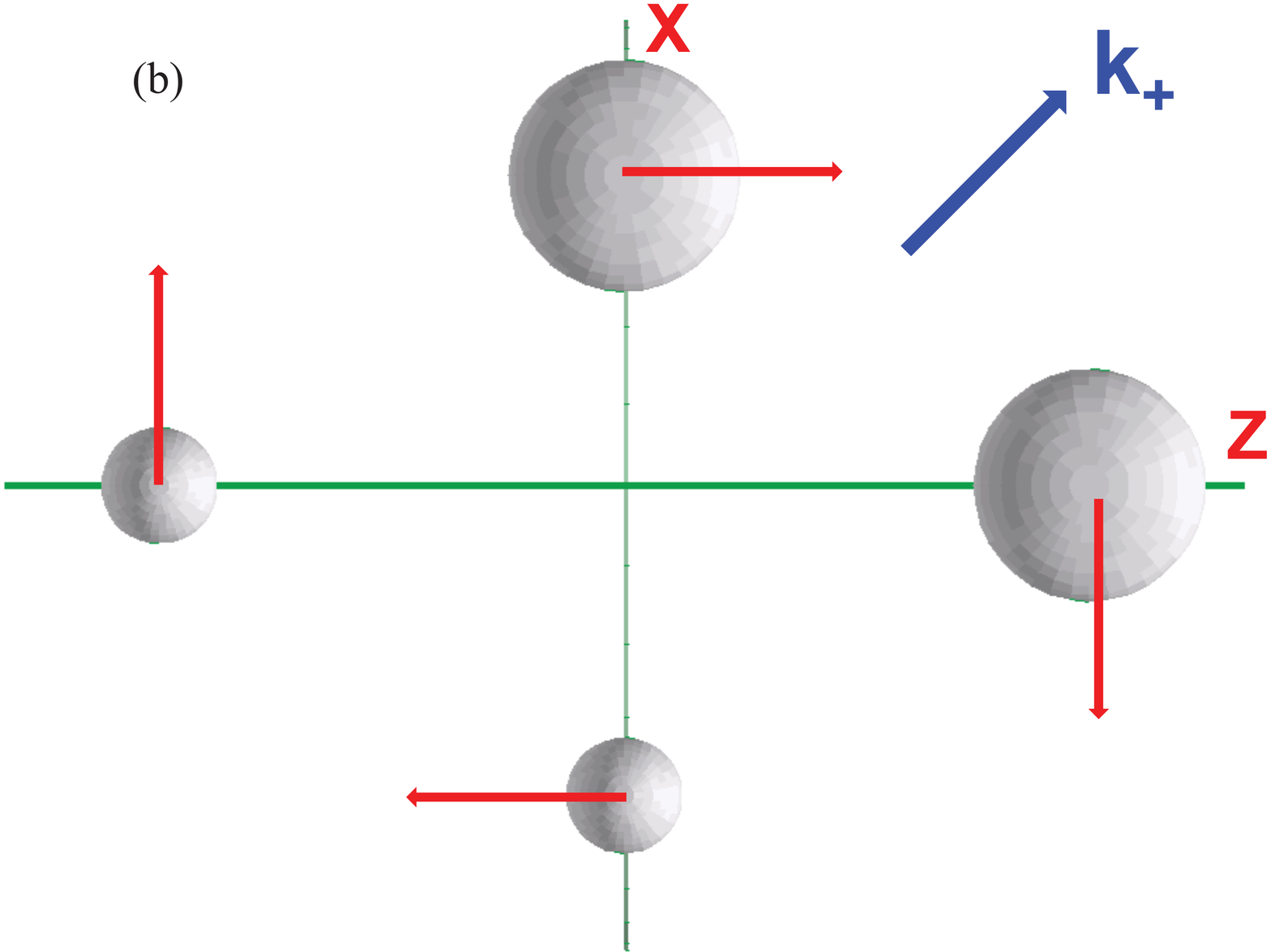}
\end{center}
\end{minipage}
\vskip -2mm
\centering
\caption{ (color online)
Four moving sources in the reaction ($[x-z]$) plane,
separated in the $x-$ and $z-$ directions. 
The sources are moving in the
directions indicated by the (red) arrows. The "tilted" detector
directions are indicated by the (blue) arrows. In the two configurations,
(a) and (b) the detector directions are different and the weights of the
sources are also different, so that the sources closer to the detector
direction have larger weights. }
\label{F-11}
\end{figure*}
Two examples on different detector configurations are given in 
Figs. \ref{F-10} and \ref{F-11}. 
We use the same equations as in the two 
source model, Eqs. (\ref{kcomp}) and (\ref{qcomp}).
A source with a larger weight factor is closer to the detector, so that
 $\omega_a$, $\omega_b$, $\omega_c$, $\omega_d$ 
correspond to 
$\vec x_s \equiv (r_x, r_z) = (d_x, 0), (-d_x, 0), (0, d_z), (0, -d_z)$
respectively. 

In case if the detector has a wide pseudorapidity acceptance, then
$\vec k_\pm$ can deviate significantly from
$k_x$, i.e. $b \ge a$ and then the weights 
are maximal for the two sources closest to $\vec k_+$ or $\vec k_-$
as indicated in Fig. \ref{F-11}.
For four sources we can use similar approach to find 
the difference of the forward and backward correlation functions.
We will use that $d_x = d_z$, $v_x = v_z$ and $a = b$.

\begin{figure}[ht] 
\begin{center}
     \includegraphics[width=6.0cm]{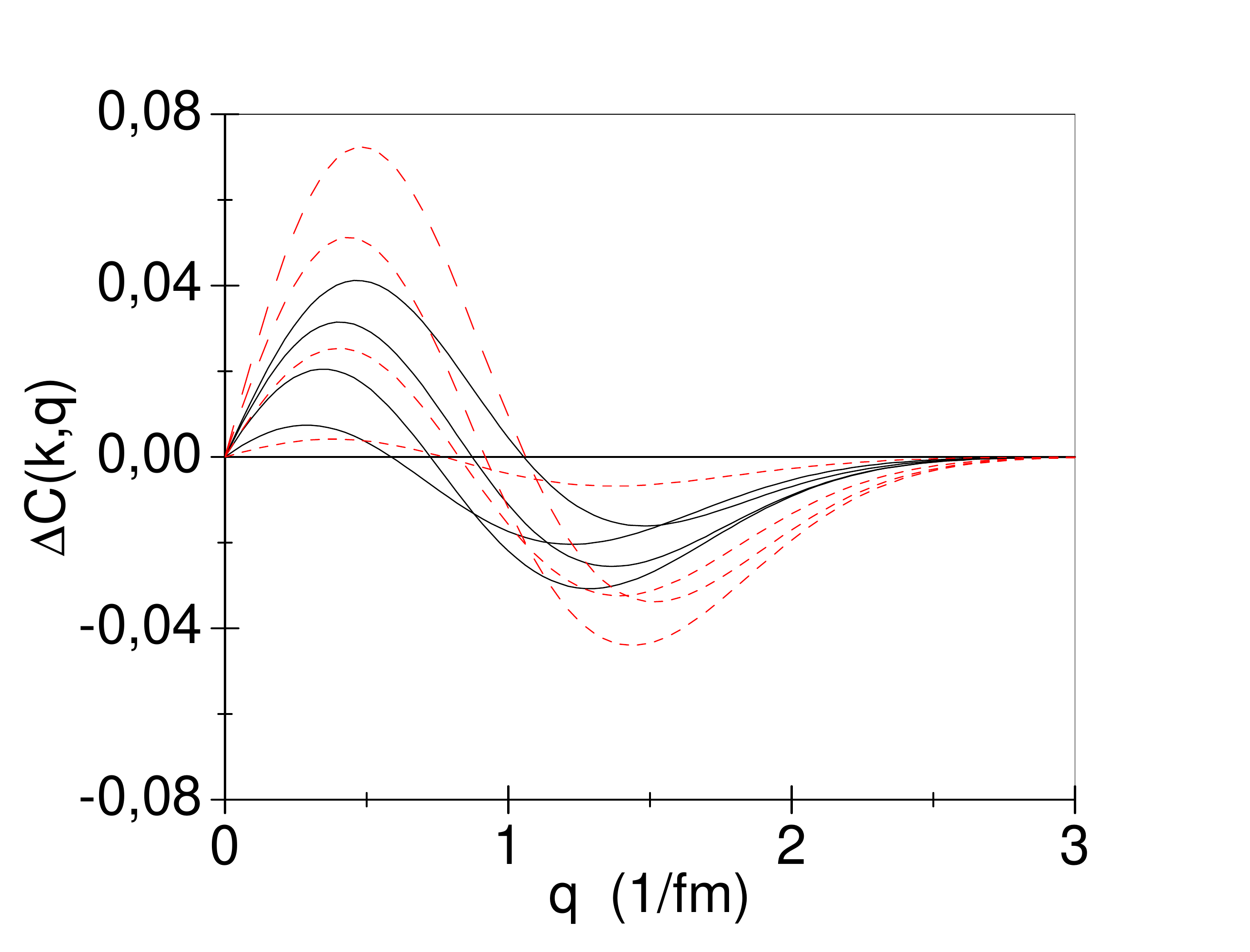}
     \includegraphics[width=6.0cm]{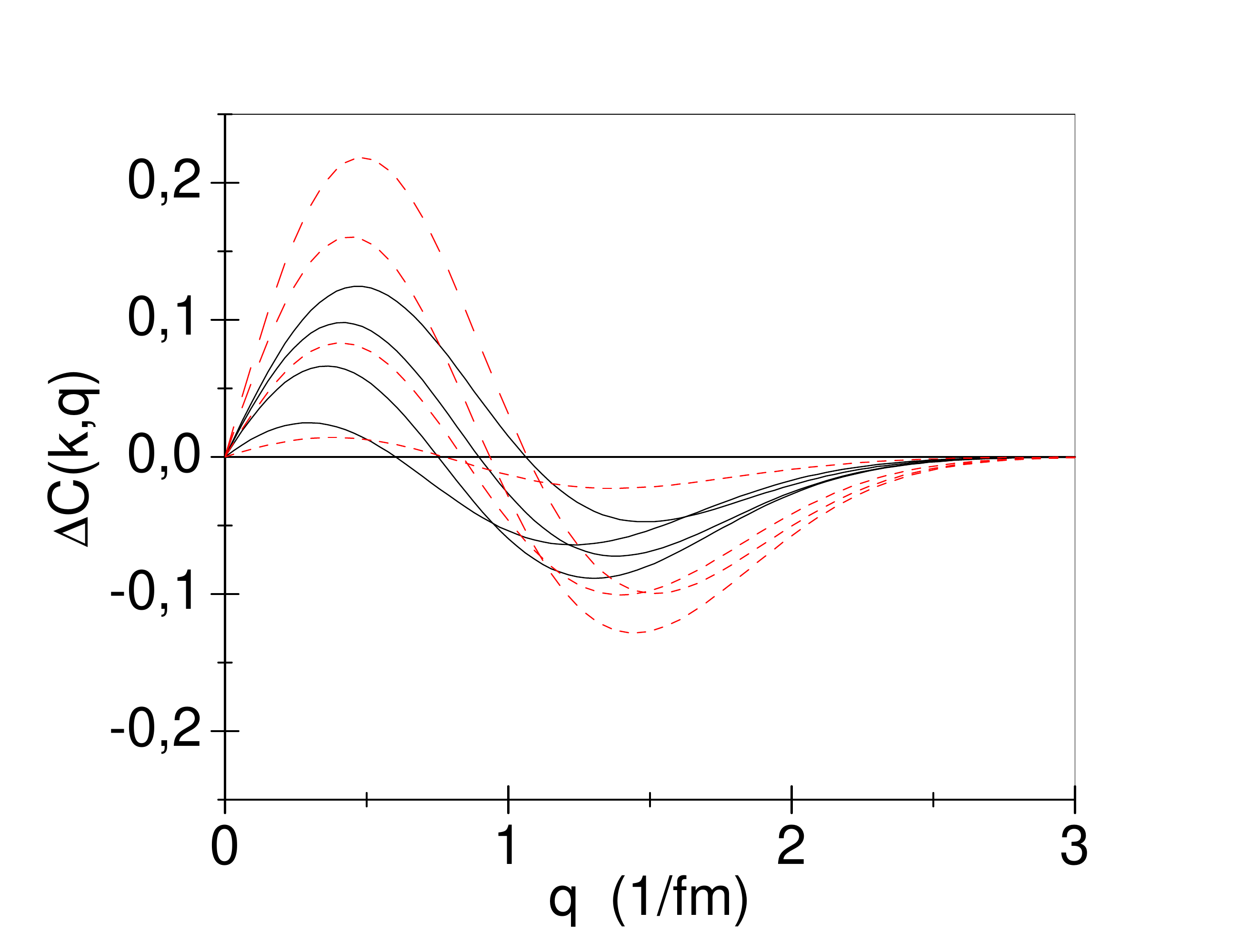}
\end{center}
\vskip -6mm
\caption{ (color online) 
{\bf Left}: The Differential Correlation Function for the weight factors:
$\omega_a = 1.25$, $\omega_b = 0.75$,
$\omega_c = \omega_d = 1.00$
for sources placed at $+x$, $-x$, $+z$ and $-z$ respectively. 
This weight distribution corresponds to the configuration
shown in Fig. \ref{F-10}.
The {\it solid black lines} are for the velocity $v_z=0.5$\,c,
and the {\it dashed red lines} are for the velocity $v_z=0.7$\,c.
Displacement is $d_x = d_z = 1.0$ fm, $T_s = 0.139$ GeV 
and $a = b = 1/\sqrt{2}$.
The values of $k$ are: for the {\it solid black lines} 
0.10, 0.50, 1.00, and 2.00 fm$^{-1}$ and
for the {\it dashed red lines}
0.10, 0.50, 1.00, and 2.00 fm$^{-1}$.
The difference is larger for smaller values of $k$.
{\bf Right}:  The Differential Correlation Function for the weight factors:
$\omega_a = 1.25$, $\omega_b = 0.75$,
$\omega_c = 1.50$, $\omega_d = 0.50$
for sources placed at $+x$, $-x$, $+z$ and $-z$ respectively
(Fig. \ref{F-11}b)
and 
$\omega_a = 1.25$, $\omega_b = 0.75$,
$\omega_c = 0.50$, $\omega_d = 1.50$
for sources placed at $+x$, $-x$, $+z$ and $-z$ respectively 
(Fig. \ref{F-11}a).
All other parameters are the same as in the left figure.
The shape is similar to the figure on the left, however the 
amplitude is different. The larger amplitude arises from the
enhanced weight factor for those sources which are closer to 
the average emission directions $\vec k_{-},\ \vec k_{+}$ in the 
Differential Correlation Function.}
\label{F-12}
\end{figure}

Some examples for the differential correlation functions are shown
in Figs. \ref{F-12}. Here due to the simplified few source
model we specified the weight distribution among the sources in a simplified
way. For realistic high resolution fluid dynamical model calculations
the realistic evaluation of emission probabilities is performed in Ref.~\refcite{CVW2014}. 
We can compare Fig. \ref{F-12} (Left) with the previously shown
two source model, Fig. \ref{F-9}, and
we see that the amplitudes are similar but the shapes are different.
First of all the sensitivity on the direction of rotation 
remained the same as in the simpler two source model. The two
extra sources, $c$ and $d$ lead to higher amplitude for the 
Differential Correlation Function, while the regular positions
of the locations of the zero points are varying due to more
sources with different weight parameters. 

If the detector acceptance is wider, then the two detectors
can be placed at more different angles. This configuration
makes the forward and backward placed sources more
accessible to the forward and backward detectors, respectively.
This is taken into account in the emission weights of our
sources. These weights are now different for the two components
of the DCF!
The result shows the tendency that the DCF
has a similar structure in the 
two source model and the four source model in a resembling
configuration.
Fig. \ref{F-12} (Right) has the same shape as Fig.  \ref{F-12} (Left), 
but the amplitude is larger.

\section{Conclusions}

In this work we study the possibility of detecting
and evaluating the rotation of a source by the specific use of
the Hanbury-Brown and Twiss method for rotating systems.
Our primary interest was the application for peripheral ultra-relativistic
heavy ions collisions with large angular momentum. In the above 
demonstrated examples the angular momentum per nucleon
is given by $\vec L_N = (1/s)\, \sum_{i=1}^{s} \vec d_i \times \vec p_i$ 
$= (\gamma m_N\,/\,s) \sum_{i=1,}^{s} \vec d_i \times \vec v_i$, where
$s$ is the number of sources. In the examples $\vec L_N$ can easily be 
obtained as $\vec d_s$ and $\vec v_s$ were always orthogonal to each other.

It turned out that it is important to take into account that
the particles reaching the detector cannot reach it with equal
probability from the near side and the far side of the emitting
object. With this fact considered we could obtain correlation
functions, which reflect the properties and also the direction
of the flow.  

We studied the Differential Hanbury Brown and Twiss
method, which made it possible to trace down the rotation
in relativistic heavy ion collisions by measuring the
correlation functions in the reaction plane at nearly transverse 
angles to the beam direction. 
The method is promising and can be performed in most
heavy ion experiments without difficulties, as well as it can be
implemented in different reaction models, like fluid dynamical
models, microscopic transport models and hybrid models.
In full scale theoretical models, the emission probabilities 
from the FO layer have been studied earlier \cite{CVW2014}.
In this case we applied the method to a high resolution,
3+1D, computational fluid dynamics model, which was used earlier 
to predict rotation, KHI, flow vorticity, and polarization 
\cite{hydro1,hydro2,CMW12,BCW2013}.

In our simple analytic models we could show that if 
we change the direction of rotation to the opposite
the Differential Correlation Function changes sign due
to the $\sinh$ function in the nominator. In this configuration
with the change of the tilt of the detector directions
we can adjust the DCF,
to the threshold value where the $\Delta C(k_\pm,q_{out})$,
is still positive, which provides a sensitive estimate
for the rotation velocity at Freeze Out.

These analytic results provide deeper insight to the methods of 
studying rotation in highly energetic systems. 
Several aspects of the sensitivity and observability 
can be discussed based on these analytic results which 
are not easily accessible in a fully realistic and complex reaction model.

\section*{Acknowledgements}

Enlightening discussions with 
Marcus Bleicher, Tam\'as Cs\"org\H o, Dariusz Miskowiec, 
Horst St\"ocker, Dujuan Wang, 
and scientists of the Frankfurt Institute for 
Advanced Studies are gratefully acknowledged.


\begin{thebibliography}{0}

\bibitem{hydro1}
  L.P. Csernai, V.K. Magas, H. St\"ocker, and D.D. Strottman,
  Phys. Rev. C {\bf 84},  024914 (2011).

\bibitem{hydro2}
  L.P. Csernai, D.D. Strottman and Cs. Anderlik,
  Phys. Rev. C {\bf 85}, 054901 (2012).

\bibitem{KHI-Wang}
   D.J. Wang, Z. N\'eda, and L.P. Csernai
   Phys. Rev. C {\bf 87}, 024908 (2013).

\bibitem{CMW12}
   L.P. Csernai, V.K. Magas, and D.J. Wang,
   Phys. Rev. C {\bf 87}, 034906 (2013). 


\bibitem{FW11-1}
S. Floerchinger and U. A. Wiedemann, 
Journal of High Energy Physics, JHEP {\bf 11}, 100 (2011).

\bibitem{FW11-2}
S. Floerchinger and U. A. Wiedemann, 
J. Phys. G: Nucl. Part. Phys. {\bf 38}, 124171 (2011).




\bibitem{StarHBT-1}
    R. Hanbury Brown and R.Q. Twiss,
    Phil. Mag. {\bf 45}, 663 (1954). 
  
\bibitem{StarHBT-2}   
    R. Hanbury Brown and R.Q. Twiss,
    Nature, {\bf 178}, 1046 (1956).

\bibitem{FirstHBTs}
   G. Goldhaber, S. Goldhaber, W. Lee and A. Pais, 
   Phys. Rev. {\bf 120}, 300 (1960).

\bibitem{DM95}
   D. Miskowiec, and E877 Collaboration, Nucl. Phys. A {\bf 590}, 557c (1995). 

\bibitem{MLisa-1}
   M.A. Lisa, N.N. Ajitanand, J.M. Alexander, et al.,
   Phys. Lett. B {\bf 496}, 1 (2000).
   
\bibitem{MLisa-2}
   M.A. Lisa, U. Heinz, U.A. Wiedemann,
   Phys. Lett. B {\bf 489}, 287 (2000).
  
\bibitem{MLisa-3}
   E. Mount, G. Graef, M. Mitrovski, M. Bleicher, M.A. Lisa, 
   Phys. Rev. C {\bf 84},  014908 (2011).

\bibitem{Pratt}
  S. Pratt, Phys. Rev. D {\bf 33}, 1314 (1986). 


\bibitem{Sin89}
  Yu.M. Sinyukov, Nucl. Phys. A498, 151c (1989).

\bibitem{QfLi07-9-1}
Qingfeng Li, J. Steinheimer, H. Petersen, M. Bleicher, 
H. St\"ocker, Phys. Lett. B {\bf 674}, 111 (2009).

\bibitem{QfLi07-9-2}
Qingfeng Li, M. Bleicher, H. St\"ocker, Phys. Lett. B {\bf 659} 525 (2008).

\bibitem{QfLi07-9-3}
Qingfeng Li, M. Bleicher, Xianglei Zhu, H. St\"oecker, 
J. Phys. G {\bf 33} 537 (2007).

\bibitem{McInnes}
  B. McInnes, 
  arXiv: 1403.3258 [hep-th]

\bibitem{BCW2013}
   F. Becattini, L.P. Csernai, D.J. Wang,
   Phys. Rev. C {\bf88}, 034905 (2013)
    

\bibitem{CVW2014}
  L. P. Csernai, S. Velle, and D. J. Wang ,
  Phys. Rev. C {\bf 89},  034916 (2014).


\bibitem{Eyyubova-1}
  L.P. Csernai, G. Eyyubova, V.K. Magas,
  Phys. Rev. C {\bf 86}, 024912 (2012). 
  
\bibitem{Eyyubova-2}
  L.P. Csernai, G. Eyyubova, V.K. Magas,
  Phys. Rev. C {\bf 86}, 019902 (2013).

\bibitem{Vovchenko}
  V. Vovchenko, D. Anchishkin, L.P. Csernai,
  Phys. Rev. C {\bf 88}, 014901 (2013).

\bibitem{Zschocke}
  S. Zschocke, Sz. Horvat, I.N. Mishustin, L.P. Csernai
  Phys. Rev. C {\bf 83}, 044903 (2011).

\bibitem{Juttner}
  F. J\"uttner, Ann. Phys. und Chemie, {\bf34} (1911) 856.

\bibitem{WF10}
  W. Florkowski: {\it Phenomenology of Ultra-relativistic
  heavy-Ion Collisions}, World Scientific Publishing Co., Singapore (2010).

\bibitem{Sinyukov-1}
A.N. Makhlin, Yu.M. Sinyukov, Z. Phys. C {\bf 39}, 69-73 (1988).

\bibitem{CF}
F. Cooper, G. Frye, Phys. Rev. D {\bf 10}, 186 (1974).

\bibitem{Cso-5}
T. Cs\"org\H{o}, Heavy Ion Phys. {\bf 15}, 1-80, (2002).






\bibitem{Horvat}  
  Sz. Horv\'at, V.K. Magas, D.D. Strottman, L.P. Csernai,
  Phys. Lett. B {\bf 692}, 277 (2010).

\bibitem{Kovtun} 
 P.K. Kovtun, D.T. Son and A.O. Starinets, 
Phys. Rev. Lett. {\bf 94}, 111601 (2005).

\bibitem{CsKM}
  L.P.~Csernai, J.I.~Kapusta, L.D.~{McLerran},
  Phys.~Rev.~Lett.~{\bf 97},  152303--4 (2006).



\bibitem{M-2}
E. Moln\'ar, L. P. Csernai, V. K. Magas, Zs. I. Lazar, A. Nyiri, and 
K. Tamosiunas, J. Phys. G {\bf 34}, 1901 (2007). 


\bibitem{M-3}
E. Moln\'ar, L. P. Csernai, V. K. Magas, A. Nyiri, and K. Tamosiunas, 
Phys. Rev. C {\bf 74}, 024907 (2006).

\bibitem{M-4}
E. Moln\'ar, L.P. Csernai, V.K. Magas, 
Acta Phys. Hung. A {\bf 27}, 359 (2006).


\bibitem{Si89-1}
Yu.M. Sinyukov, Yad. Fiz. {\bf 50}, 228 (1989).

\bibitem{Si89-2}
Yu.M. Sinyukov, Sov. J. Nucl. Phys.
{\bf 50}, 143 (1989).

\bibitem{Si89-3}
Yu.M. Sinyukov, Z. Phys. C {\bf 43}, 401 (1989).

\bibitem{Bugaev}
K.A. Bugaev, Nucl. Phys. A {\bf 606}, 559 (1996).

\bibitem{ALM99}
      Cs. Anderlik, Z.I. L\'az\'ar, V.K. Magas, L.P. Csernai, H. St\"ocker
      and W. Greiner, Phys. Rev. C {\bf 59},  388 (1999).




\bibitem{ACG99}
      C. Anderlik, L.P. Csernai, F. Grassi, W. Greiner,
      Y. Hama, T. Kodama, Z.I. L\'az\'ar, V.K. Magas and H. St{\"o}cker,
      Phys. Rev. C {\bf 59}, 3309 (1999) .
      
\bibitem{MAC99}
      V.K. Magas, C. Anderlik, L.P. Csernai, F. Grassi, W. Greiner,
      Y. Hama, T. Kodama,  Z.I. L\'az\'ar and H. St{\"o}cker,
      Nucl. Phys. A {\bf 661}, 596c (1999). 

\bibitem{Cs02}
      L.P. Csernai, J. Phys. G {\bf 28}, 1993 (2002).

\bibitem{MAA03}
      V.K. Magas, A. Anderlik, Cs. Anderlik and L.P. Csernai,
      Eur. Phys. J. C {\bf 30}, 255 (2003).

\bibitem{TC04}
      K. Tamosiunas and L.P. Csernai, 
      Eur. Phys. J. A {\bf 20}, 269 (2004).

\bibitem{MCM05}
      V.K. Magas, L.P. Csernai, E. Moln\'ar, A. Nyiri, K. Tamosiunas,
      Nucl. Phys. A {\bf{749}}, 202 (2005).

\bibitem{MCM06b}
   V.K. Magas, L.P. Csernai, E. Moln\'ar,
   Acta Phys. Hung. A {\bf{27}}, 351 (2006).

\end{thebibliography}
\end{document}